\documentclass[aps,prd,12pt,superscriptaddress,nofootinbib,floatfix]{revtex4-2} 
\usepackage{epsfig}
\usepackage{graphicx}
\usepackage[normalem]{ulem}
\usepackage{amssymb}
\usepackage{amsmath}
\usepackage{amsfonts}
\usepackage{amsthm}
\usepackage{latexsym}
\usepackage{verbatim}
\usepackage{mathtools}
\usepackage{setspace}
\usepackage{slashed}
\usepackage[all]{xypic}
\usepackage{psfrag}
\usepackage{comment}
\usepackage{array}
\usepackage{float}
\usepackage{orcidlink}
\usepackage{textcomp}
\usepackage{wasysym,subfigure,bm,mathrsfs,lipsum}
\usepackage{hyperref}
\hypersetup{
	colorlinks=true,
	linkcolor=red,
	citecolor=blue,
}

\begin{document}
\title{Rescattering of non-minimal coupling scalar particles during inflation}
\author{Zhe Yu}
\email{yuzhe@nbu.edu.cn}
\affiliation{Institute of Fundamental Physics and Quantum Technology,
 Department of Physics, School of Physical Science and Technology,
 Ningbo University, Ningbo, Zhejiang 315211, China}
\author{Xunliang Yang}
\email{yangxunliang22@mails.ucas.ac.cn}
\affiliation{School of Fundamental Physics and Mathematical Sciences, Hangzhou Institute for Advanced Study, UCAS, Hangzhou 310024, China }
\affiliation{ Institute of Theoretical Physics, Chinese Academy of Sciences,Beijing 100190,China  }
\affiliation{University of Chinese Academy of Sciences, Beijing 100049, China}
\author{Yungui Gong\orcidlink{0000-0001-5065-2259}}
\email{gongyungui@nbu.edu.cn}
\affiliation{Institute of Fundamental Physics and Quantum Technology,
 Department of Physics, School of Physical Science and Technology,
 Ningbo University, Ningbo, Zhejiang 315211, China}

\begin{abstract}
We investigate the rescattering effects arising from non-minimally coupled scalar particles $\chi$ that are suddenly produced during inflation. 
The coupling term $\xi R \chi^2$ significantly enhances resonant particle production compared to minimal coupling scenarios. 
Consequently, the produced $\chi$ particles rescattering off the homogeneous inflaton condensate $\phi$, 
generating abundant $\delta\phi$ quanta within very short time intervals. 
This process leads to characteristic enhancements in the power spectrum of primordial curvature perturbations at scales corresponding to the moments of particle production. 
When this occurs at small scales, the power spectrum amplitude can reach as high as $\mathcal{O}(10^{-2})$. 
Furthermore, analysis of the equilateral bispectrum shows that this mechanism also induces substantial non-Gaussian features.
\end{abstract}

\maketitle

\section{Introduction}
\label{sec:1}
The inflationary paradigm \cite{Guth:1980zm,Starobinsky:1980te,Sato:1980yn,Linde:1981mu} resolves fundamental issues of the standard cosmological model and has become the prevailing framework for describing the early universe. 
During inflation, the accelerated expansion causes the comoving Hubble radius to shrink exponentially, 
stretching quantum vacuum fluctuations beyond the horizon where they freeze \cite{Mukhanov:1981xt,Sasaki:1983kd,Kodama:1984ziu}. 
Among these perturbations, the primordial curvature perturbation which is associated with the scalar modes, dominates the observed anisotropies in the cosmic microwave background (CMB).
Precision measurements from contemporary CMB experiments such as Planck \cite{Planck:2018jri,Planck:2018vyg}, BICEP/Keck \cite{BICEP2:2018kqh,BICEP:2021xfz}, and the latest ACT results \cite{ACT:2025tim,ACT:2025fju} have tightly constrained the curvature perturbations. 
These datasets determine the amplitude of the power spectrum at the pivot scale $k=0.05\rm{Mpc}^{-1}$ to be approximately $2.1\times 10^{-9}$ with slight scale dependence \cite{Planck:2018jri}, 
constrain the scalar spectral index to $n_s=0.974\pm 0.03$ through joint analyses~\cite{ACT:2025tim}, 
and set an upper limit on the tensor-to-scalar ratio of $r<0.036$ (95\% CL),
thereby limiting the presence of primordial gravitational waves (GWs) \cite{BICEP:2021xfz}. While these observations precisely characterize curvature perturbations on large (CMB) scales, their properties on small scales remain largely unconstrained due to observational limitations. This gap motivates extensive theoretical investigations into mechanisms that can modify primordial curvature perturbations from standard slow-roll expectations at these small scales~\cite{Cai:2021wzd,Peng:2021zon,Fu:2022ypp,Dimopoulos:2017ged}.
In this work, we explore how rescattering of non-minimally coupled scalar particles leads to corrections in the power spectrum of primordial curvature perturbations and induces distinctive equilateral-type bispectra.

Scalar particle production during inflation has been widely studied~\cite{Chung:1998bt,Chung:2018ayg,Ford:2021syk,Kolb:2023ydq}, with particular attention to burst-like scenarios characterized by rapid particle generation~\cite{Chung:1999ve,Romano:2008rr,Battefeld:2010sw,Cook:2011hg,Fedderke:2014ura,Kim:2021ida}. 
In these scenarios, an additional scalar field $\chi$ is coupled to the inflaton $\phi$ through an interaction of the form $g^2(\phi-\phi_0)^2\chi^2$. 
When the inflaton $\phi$ approaches the value  $\phi_0$, 
the system experiences a non-adiabatic transition, violating the adiabatic condition $\dot{\omega}_k\leq \omega^2_k$, where $\omega_k^2\approx g^2(\phi - \phi_0)^2$ characterizes the frequency of mode $\chi_k$. 
Efficient particle production requires the nonadiabatic time scale,  
$\Delta t=1/(-g\dot{\phi}_0)^{\frac{1}{2}}$ to be shorter than a Hubble time. 
After the burst, the effective mass of $\chi$ grows quickly, preventing condensate formation. 
Because occupation numbers satisfy $n_k < 1$ for all modes, 
this minimal model yields inefficient GW production.
To boost particle production efficiency, non-minimal couplings to gravity are introduced \cite{Li:2019ves,Clery:2022wib,Yu:2023ity,Capanelli:2024rlk,He:2024bno}. 
Specifically, the interaction $\xi R \chi^2$ induces a tachyonic instability near $\phi \approx \phi_0$, 
with the effective mass-squared given by $m_{\rm eff}^2 = g^2(\phi-\phi_0)^2 -\xi R$.
When this quantity becomes negative,
$\chi$-particle are exponentially amplified. 
However, the large occupation numbers that result lead to significant backreaction on the inflaton dynamics~\cite{Yu:2023ity}, 
which quenches further particle production. Although GW signals in this non-minimal scenario are stronger than in the minimal model, 
they remain below the sensitivity of detectors such LISA\cite{Danzmann:1997hm,LISA:2017pwj}, TianQin \cite{Luo:2015ght,Gong:2021gvw}, and Taiji \cite{Hu:2017mde}, due to rapid mass acquisition by $\chi$ particles. 
Consequently, signatures of particle production imprinted on curvature perturbations become critical observational targets. 
In our recent work \cite{Yang:2025zap}, employing the Hartree approximation, we incorporated backreaction effects from strongly produced particles via non-minimal coupling in the Starobinsky inflation model, 
demonstrating an enhancement in the scalar spectral index $n_s$ within two e-folds consistent at the $1\sigma$ level with the current P+ACT+LB+BK18 constraints.
 
Going beyond the Hartree approximation, rescattering processes constitute the leading-order source of backreaction: 
rapidly produced $\chi$-particles interact nonlinearly with the inflaton background, 
transferring momentum to the inflaton fluctuations $\delta\phi$. 
These rescattering effects following sudden particle production were first analyzed by Barnaby et al. \cite{Barnaby:2009mc},
who treated the produced particles as a classical source term in the equation of motion for $\delta\phi$, 
deriving enhanced curvature perturbations. However, their classical treatment omitted important leading-order quantum contributions. 
Pearce et al. \cite{Pearce:2017bdc} revisited this problem using the in-in formalism, 
showing that the correction to the power spectrum is subdominant, with $\Delta P_{\mathcal{\zeta}}/P_{\mathcal{\zeta}} \lesssim 10^{-2}$.
More recently, Fumagalli et al. \cite{Fumagalli:2023loc} demonstrated that rescattering of amplified curvature perturbations generates dominant one-loop corrections to the power spectrum. 
Analogously, in scenarios with non-minimally coupled particle production, rescattering can amplify the power spectrum up to $\mathcal{O}(10^{-2})$ on small scales and produce significant non-Gaussianities characterized by 
$f_{\rm NL} \sim \mathcal{O}(1000)$. 
Such effects could provide seeds for primordial black holes (PBHs) and scalar-induced gravitational waves (SIGWs).

This paper is organized as follows. 
In Section \ref{sec:2}, we present the non-minimal coupling model describing sudden particle production during inflation and analyze its rescattering imprints on curvature perturbations. 
Subsection \ref{sec:21} reviews the particle production mechanism using the Bogoliubov coefficient formalism. 
Subsection \ref{sec:22} derives the leading-order corrections to the curvature power spectrum including rescattering via the in-in formalism. 
Subsection \ref{sec:23} investigates the induced non-Gaussianities by computing the equilateral bispectrum and evaluating the $f_{\rm NL}$ parameter over relevant scales. 
We conclude with a discussion in Section \ref{sec:3}. 
Throughout, we adopt natural units $c = \hbar = 1$ and set the reduced Planck mass as $M_{\mathrm{p}} \equiv (8\pi G)^{-1/2} = 1$.

\section{The rescattering effects}
\label{sec:2}
\subsection{Particle Production and backreaction}
\label{sec:21}
This subsection reviews the mechanism of particle production for a scalar field non-minimally coupled to gravity within the framework of general single-field slow-roll inflation \cite{Yu:2023ity}. 
We consider a model that includes an additional quantum scalar field 
$\chi$, 
which has no homogeneous background component initially. 
The field $\chi$ is directly coupled to the inflaton $\phi$ 
and non-minimally coupled to the Ricci scalar $R$. 
The action is given by
\begin{align}
\label{action}
S=\int d^4 x \sqrt{-\hat{g}}\left[\frac{M_{\mathrm{p}}^2}{2}R-\frac{1}{2}\nabla^\mu \phi  \nabla_\mu \phi-V(\phi)-\frac{1}{2} \nabla^\mu\chi\nabla_\mu\chi-\frac{g^2}{2}\left(\phi-\phi_0\right)^2 \chi^2+\frac{1}{2}\xi R \chi^2\right],
\end{align}
For convenience, in this paper, we assume $\dot{\phi}<0$ and $\partial_\phi V>0$. 
Varying the action \eqref{action} with respect to the field $\chi$ yields the equation of motion
\begin{align}
\label{EoM_chi}
\ddot{\chi}+3H\dot{\chi}-\frac{1}{a^2}\nabla^2\chi+\left[g^2\left(\phi-\phi_0\right)^2-\xi R\right]\chi=0.
\end{align}
The quantum field $\chi$ can be expanded in terms of creation and annihilation operators as
\begin{align}
\label{decomposition_chi}
\chi=\frac{1}{(2\pi)^{3/2}} \int d^3k\left[\chi_k\hat{a}_{\boldsymbol{k}} + \chi^\ast_{-k} \hat{a} ^\dagger_{-\boldsymbol{k}}  \right]e^{i \boldsymbol{k} \cdot \boldsymbol{x}},
\end{align}
where the annihilation and creation operators $\hat{a}_{\boldsymbol{k}}$ and $\hat{a}^{\dagger}_{\boldsymbol{k}}$ satisfy the canonical commutation relation $[\hat{a}_{\boldsymbol{k}},\hat{a}^\dagger_{\boldsymbol{k}^\prime}]=\delta(\boldsymbol{k}-\boldsymbol{k}^\prime)$. 
Substituting the mode decomposition \eqref{decomposition_chi} into Eq. \eqref{EoM_chi}, 
the mode functions $\chi_k$ obey
\begin{align}
\label{EoM_chi_k}
\ddot{\chi}_k+3H \dot{\chi}_k+\omega_k^2 \chi_k =0,
\end{align}
with
\begin{align}
\label{omega_1}
\omega_k^2&=\frac{k^2}{a^2}+g^2\left(\phi-\phi_0\right)^2-6\xi(\dot H + 2H^2)\nonumber\\& \simeq \frac{k^2}{a^2}+g^2\left(\phi-\phi_0\right)^2-12\xi H^2,
\end{align}
Here we used the relation for the Ricci scalar in a Friedmann–Lemaître–Robertson–Walker (FLRW) background, $R=6(\dot{H} + 2H^{2})$,
assuming the slow-roll parameter $\epsilon = -\dot{H}/H^{2} \ll 1$.
To obtain an analytic solution for $\chi_k$,
we first neglect the backreaction of $\chi$.
Defining a rescaled field $\hat{\chi}_k=a\chi_k$,
Eq.~\eqref{EoM_chi_k} becomes
\begin{align}
\label{Conformal_EoM_Chi}
\hat{\chi}^{\prime\prime}_k+\hat{\omega}_k^2 \hat{\chi}_k =0,
\end{align}
where the prime denotes derivative with respect to conformal time $\tau$,
and
\begin{align}
\hat{\omega}_k^2=k^2+g^2\left(\phi-\phi_0\right)^2a^2-2(6\xi+1)H^2a^2.
\end{align}
To ensure efficient particle production, we assume that the production duration is shorter than a Hubble time. 
Therefore, near the point $\phi=\phi_0$, 
the inflaton’s behavior can be approximated as linear, $\phi=\phi_0+\dot{\phi}_0(t-t_0)\approx\phi_0-\dot{\phi}_0(\tau/\tau_0-1)/H$, 
where the Hubble parameter $H$ is assumed constant during inflation, and quantities with the subscript $0$ are evaluated at $\phi=\phi_0$. 
An approximate upper bound on the particle production duration, 
defined as $\Delta t \equiv t_e - t_i \simeq 2(t_e-t_0)$ with $t_i$ and $t_e$ denoting the start and end times of production, 
can be derived by requiring that the effective frequency squared for modes with negligible momentum satisfies $\hat{\omega}_k^2 < 0$ when $k\simeq 0$. 
This leads to the constraint
\begin{align}
\label{time}
\gamma \equiv \frac{H \Delta t}{2} =\frac{H^2 \sqrt{2(6 \xi+1)}}{-g \dot{\phi}_0}<\frac{1}{2}.
\end{align}
During the particle production phase, the expansion of the universe can be neglected, 
allowing us to approximate the scale factor as roughly constant, $a\approx a_0$.
Under this approximation, Eq. \eqref{Conformal_EoM_Chi} reduces to the Weber equation. 
Assuming the initial state is the Bunch-Davies vacuum, the exact solution for the mode functions takes the form   
\begin{align}
\label{Exact_Sol}
\hat{\chi}_k=2^{-\frac{3}{4}}\left[\frac{1}{\sqrt{\sigma_k}} W\left(-\frac{\kappa^2}{2} ;-\sqrt{2} y\right)+i \sqrt{\sigma_k} W\left(-\frac{\kappa^2}{2} ; \sqrt{2} y\right)\right],
\end{align}
where $W(a;x)$ denotes the parabolic cylinder function, and
\begin{align}
y=\left(\frac{\tau}{\tau_0}-1\right) \frac{\sqrt{-g \dot{\phi}_0}}{H}, \, \kappa^2=-\frac{k^2 \tau_0^2 H^2}{g \dot{\phi}_0}+\frac{2(1+6 \xi) H^2}{g \dot{\phi}_0},\, \sigma_k=\sqrt{1+e^{-\pi \kappa^2}}-e^{-\frac{\pi \kappa^2}{2}}.
\end{align}
For times later than $t_0+\Delta t/2$, 
the system returns to an adiabatic regime, 
and the solution can be expressed as a linear combination of incident and outgoing waves
\begin{align}
\label{Adiabatic_Exact_Sol}
\hat{\chi}_k=\alpha_k \Phi_k+\beta_k \Phi_k^\ast,   
\end{align}
where the Bogoliubov coefficients $\alpha_k=i\left(\sigma_k+1/\sigma_k\right)/2$ and $\beta_k=i\left(\sigma_k-1/\sigma_k\right)/2$ satisfy $|\alpha_k|^2-|\beta_k|^2=1$. 
The adiabatic mode functions are $\Phi_k=\exp \left[-i \int \hat{\omega}_k d\tau\right]/\sqrt{2 \hat{\omega}_k}$, where the mode frequency can be approximated as
\begin{align}
\label{omega_3}
 \hat{\omega}_k\approx ag\left(\phi_0-\phi\right).   
\end{align}
Using these Bogoliubov coefficients, the comoving particle occupation number is obtained as
\begin{align}
\label{beta}
\left|\beta_k\right|^2=e^{-\frac{\pi\left(k \tau_0\right)^2 H^2}{-g \dot{\phi}_0}} e^{\frac{2 \pi(1+6 \xi) H^2}{-g \dot{\phi}_0}},    
\end{align}
From the expression for the occupation number \eqref{beta}, we see that as the parameter $\xi$ grows large, the particle number diverges. 
This signals the necessity to include backreaction effects on the background dynamics to constrain $\xi$.
Instead of solving the full coupled equations, such backreaction can be effectively incorporated by modifying the inflaton potential to an effective potential, 
$V_{\rm{eff}}=V+g^2(\phi-\phi_0)^2\left\langle\chi^2\right\rangle$. 
When the inflaton field $\phi$ crosses $\phi_0$, 
excessive particle production can invert the sign of the derivative $dV_{\rm eff}/d\phi$, 
thereby creating a potential barrier that prematurely terminates inflation. 
To guarantee that slow-roll inflation ends as originally intended, the following inequality must hold,
\begin{align}
\frac{dV_{\rm eff}}{d\phi} =g^2\left(\phi-\phi_0\right)\left\langle\chi^2\right\rangle + \frac{d V}{d \phi} \geq 0.
\end{align}
Using Eq.~\eqref{beta}, the backreaction term evaluates approximately to 
\begin{align}
g^2\left(\phi-\phi_0\right)\left\langle\chi^2\right\rangle\approx-\frac{g}{a^3} \int \frac{d k}{2 \pi^2} k^2\left|\beta_k\right|^2=-e^{\frac{2 \pi(1+6 \xi) H^2}{-g \dot{\phi}_0}} \frac{g}{16 \pi^3 a^3} \frac{(-g \dot{\phi}_0)^{\frac{3}{2}}}{H^3 \tau_0^3}.
\end{align}
From this condition, we derive an upper bound on $\xi$
\begin{align}\label{Constrain_xi}
\exp\left[{\frac{2 \pi(1+6 \xi) H^2}{-g \dot{\phi}_0}}\right]=\frac{d V}{d \phi} \frac{16 \pi^3}{g(-g \dot{\phi}_0)^{\frac{3}{2}}}.    
\end{align}

\subsection{Rescattering}
\label{sec:22}
In this subsection, we investigate the  primordial curvature perturbation by incorporating the effect of leading-order rescattering of the particle $\chi$ within the framework of the in-in formalism. 
In this approach, the expectation value of an operator $Q$ at time $t$ is given by
\begin{align}
\label{inin}
\left\langle Q\right\rangle=\left\langle0|U^\dagger(t) Q U(t)  |0\right\rangle , 
\end{align}
where $Q$ on the right-hand side denotes the operator in the interaction picture (free field theory), termed the unperturbed operator. 
The time-evolution operator $U(t)$ is defined as
\begin{align}
U(t)=T\exp\left(-i\int^t dt^\prime H_I(t^\prime) \right),
\end{align}
with the interaction Hamiltonian  $H_I(t)$ and $T$ is the time-ordering operator. 
As mentioned in the introduction, the particle 
$\chi$ transfers momentum to produce quanta $\delta\phi$. 
Accordingly, it is natural to decompose the inflaton field as $\phi_{\mathbf{tot}}=\phi+\delta\phi$,  
where $\phi$ is the homogeneous background inflaton. 
Since we focus on leading-order corrections to the power spectrum of the curvature perturbation,
the quantum fluctuation $\delta\phi$ is treated as the first-order perturbation.  
For consistency, we adopt the first-order metric perturbations in the spatially flat gauge
\begin{align}
ds^2=-(1+2\Psi)dt^2+2a\partial_iBdtdx^i+a^2\delta_{ij} dx^idx^j. 
\end{align}
To apply Eq. \eqref{inin},  we need the unperturbed mode functions of $\delta\phi$ which correspond to standard vacuum fluctuations. 
These obey the Sasaki–Mukhanov equation, 
\begin{equation}
\label{eq_sasaki}
\delta \ddot{\phi}_k+3H\delta\dot{\phi}_k+\left(\frac{k^2}{a^2}+M_{\mathbf{eff}}^2\right)\delta\phi_k=0,   
\end{equation}
where the effective mass is
\begin{align}
M_{\mathbf{eff}}^2=\frac{\partial^2V}{\partial\phi^2}-\frac{1}{a^3}\frac{d}{dt}\left(\frac{a^3}{H}\dot{\phi}^2 \right)\nonumber.
\end{align}
Under the slow-roll approximation, this reduces to $M_{\mathrm{eff}}^2 \approx H^2(3\eta - 6\epsilon)$, with the second slow-roll parameter defined as $\eta \equiv \partial^2 V / \partial \phi^2$. 
Following Ref.~\cite{Pearce:2017bdc}, 
we neglect slow-roll corrections to isolate the power spectrum features arising solely from particle production. 
Under this assumption, the solution to Eq.~\eqref{eq_sasaki} is
\begin{align}
 \delta\phi_k=\frac{H}{\sqrt{2k}}\left(-\tau+\frac{i}{k}\right)e^{-ik\tau} ,
\end{align}
From this mode function, the unperturbed power spectrum of the primordial curvature perturbation is
\begin{equation}
 P_{\zeta}^{(0)}=\frac{H^4}{4\pi^2\dot{\phi}^2}.   
\end{equation}

\begin{figure}
\centering
\includegraphics[width=0.25\textheight]{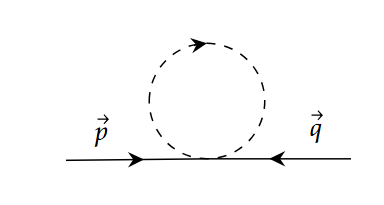}
\includegraphics[width=0.25\textheight]{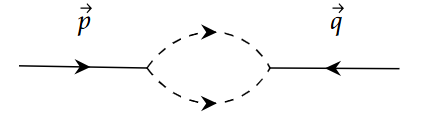}
\caption{The Feynman diagram for the leading order processes for the rescaterring of particle $\chi$. The left and right panels correspond to Eq.~\eqref{loop1} and Eq.~\eqref{loop2}, respectively.  }
\label{fig1}
\end{figure}

Next, we determine the dominant interaction terms to construct the evolution operator. 
As argued in the previous subsection, when the non-minimal coupling coefficient $\xi$ is sufficiently constrained, 
the backreaction on the background remains subdominant, and slow-roll conditions persist. 
Accordingly, the leading-order interactions couple directly to $\delta\phi$.
Details of the derivation are presented in Appendix
\ref{A}, and the resulting interaction Hamiltonian in conformal time reads 
\begin{align}
\label{Intaction_final}
H_{I}=\int dx^3 a^4 \left[g^2\left(\phi-\phi_0\right)\chi^2\delta\phi+\frac{1}{2}g^2\chi^2\delta\phi^2\right].
\end{align}
The cubic and quartic interactions in Eq. \eqref{Intaction_final} generate one-loop corrections to the curvature perturbation. 
By substituting $Q=\delta\phi_{\boldsymbol{p}}\delta\phi_{\boldsymbol{q}}$ into Eq. \eqref{inin}, 
one obtains two types of correlators for  $\delta\phi$,
\begin{align}
\label{loop1}
\left\langle\delta \phi_{\boldsymbol{p}}(\tau) \delta \phi_{\boldsymbol{q}}(\tau)\right\rangle_{1}&=\left\langle i\int^{\tau}_{\tau_*} d \tau^{\prime} H_I^{(4)}(\tau^\prime)\delta \phi_{\boldsymbol{p}}(\tau)\delta \phi_{\boldsymbol{q}}(\tau)-i\int^{\tau}_{\tau_*} d \tau^{\prime}\delta \phi_{\boldsymbol{p}}(\tau)\delta \phi_{\boldsymbol{q}}(\tau)H_I^{(4)}(\tau^{\prime})\right\rangle \nonumber\\&=(2\pi)^3\delta^{(3)}\left(\boldsymbol{p}+\boldsymbol{q}\right)\left|\delta\phi_p\right|_{(1)}^2 , 
\end{align}
\begin{align}
\label{loop2}
\left\langle\delta \phi_{\boldsymbol{p}}(\tau) \delta \phi_{\boldsymbol{q}}(\tau)\right\rangle_2=&\left\langle\int^\tau_{\tau_*} d \tau^{\prime} H_I^{(3)}(\tau^{\prime}) \delta \phi_{\boldsymbol{p}}(\tau) \delta \phi_{\boldsymbol{q}}(\tau) \int^\tau_{\tau_*} d \tau^{\prime \prime} H_I^{(3)}(\tau^{\prime \prime}) \right\rangle
\nonumber \\& -   \left\langle\int^\tau_{\tau_*} d \tau^{\prime} \int^{\tau^{\prime}}_{\tau_*} d \tau^{\prime \prime}H^{(3)}(\tau^{\prime\prime}) H^{(3)}(\tau^{\prime})\delta \phi_{\boldsymbol{p}}(\tau) \delta \phi_{\boldsymbol{q}}(\tau)\right\rangle \nonumber \\&  - \left\langle\int^\tau_{\tau_*} d \tau^{\prime} \int^{\tau^{\prime}}_{\tau_*} d \tau^{\prime \prime} \delta \phi_{\boldsymbol{p}}(\tau) \delta \phi_{\boldsymbol{q}}(\tau) H^{(3)}(\tau^{\prime}) H^{(3)}(\tau^{\prime \prime})\right\rangle\nonumber\\=&(2\pi)^3\delta^{(3)}\left(\boldsymbol{p}+\boldsymbol{q}\right)\left|\delta\phi_p\right|_{(2)}^2 ,
\end{align}
where $\tau_*$ denotes the onset of particle production. 
The associated Feynman diagrams are depicted in Figure \ref{fig1}.
To evaluate the final power spectrum, we require the unperturbed propagators of
$\delta\phi_{\boldsymbol{k}}$ and the particle $\hat{\chi}_{\boldsymbol{k}}$, which are given by
\begin{align}
\label{Pro_phi}
\langle\delta\phi_{\boldsymbol{p}}(\tau_1)\delta\phi_{\boldsymbol{q}}(\tau_2)\rangle=(2\pi)^3\delta^{(3)}\left(\boldsymbol{p}+\boldsymbol{q}\right)\delta\phi_p(\tau_1)\delta\phi_p^\ast(\tau_2),    
\end{align}
\begin{align}
\label{Pro_phi}
&\langle:\hat{\chi}_{\boldsymbol{p}}(\tau_1)\hat{\chi}_{\boldsymbol{q}}(\tau_2):\rangle=(2\pi)^3\delta^{(3)}\left(\boldsymbol{p}+\boldsymbol{q}\right) f_p(\tau_1,\tau_2),
\end{align}
where the normal ordering removes vacuum contractions, and
\begin{align}
f_{p}(\tau_1,\tau_2)=&
            \alpha_p(\tau_1)\beta_p^*(\tau_2)\Phi_p(\tau_1)\Phi_p(\tau_2)+
            \beta_p(\tau_1)\beta_p^*(\tau_2)\Phi_p^*(\tau_1)\Phi_p(\tau_2)\nonumber\\&+
            \beta_p^*(\tau_1)\beta_p(\tau_2)\Phi_p(\tau_1)\Phi_p^*(\tau_2)+
            \beta_p(\tau_1)\alpha_p^*(\tau_2)\Phi_p^*(\tau_1)\Phi_p^*(\tau_2),
\end{align}
In calculating the $\chi$-propagator, UV divergences in the $\chi$-particle energy stemming from adiabatic contributions are removed via normal ordering renormalization \cite{Chung:2004nh}. 
For subsequent calculations, 
we also evaluate the retarded Green's function
\begin{align}
\label{Green_phi}
 [\delta \phi_{\boldsymbol{p}}(\tau^\prime),\delta \phi_{\boldsymbol{q}}(\tau)]&=(2\pi)^3\delta^{(3)}\left(\boldsymbol{p}+\boldsymbol{q}\right)\left(\delta \phi_p(\tau^\prime)\delta \phi_p^{*}(\tau)-\delta \phi_p^*(\tau^\prime)\delta \phi_p(\tau)\right)\nonumber\\&=i(2\pi)^3\delta^{(3)}\left(\boldsymbol{p}+\boldsymbol{q}\right) G_p(\tau^\prime,\tau).
\end{align}
Under normal ordering renormalization,
the Green's functions for $\chi$ vanish,
\begin{equation}
\label{Green_chi}
 [: \chi_{\boldsymbol{p}}(\tau^\prime),\chi_{\boldsymbol{q}}(\tau):]=0.
\end{equation}
implies that the $\chi$ field transitions effectively into a classical field after particle production. 
By combining Eqs.~\eqref{Pro_phi}–\eqref{Green_chi} with the Feynman diagrams, we derive the amplitudes for the processes,
\begin{align}
\label{u1}
\left|\delta\phi_p(\tau)\right|_{(1)}^2 =-2g^2 \int^{\tau}_{\tau_*} d\tau_1 a^2(\tau_1) \int \frac{dk^3}{(2\pi)^3} f_k(\tau_1,\tau_1) Im\left[\delta \phi_p^2(\tau_1) \delta \phi_p^{*2}(\tau)\right]
\end{align}
\begin{align}
\label{u2}
\left|\delta\phi_p(\tau)\right|_{(2)}^2\!=&\!-2g^4\!\int^\tau_{\tau_*} \!d\tau_1 d\tau_2 a^2\!(\tau_1)a^2(\tau_2)\left[\phi(\tau_1)\!-\!\phi_0\right] \!\left[\phi(\tau_2)\!-\!\phi_0\right] \nonumber \\ &\! \times\int \!\frac{dk^3}{(2\pi)^3} G_{p}(\tau_2,\tau)G_{p}(\tau_1,\tau)f_k(\tau_1,\tau_2)f_{|\boldsymbol{k}-\boldsymbol{p}|}(\tau_1,\tau_2).
\end{align}
We now proceed to compute the power spectrum of the primordial curvature perturbation. 
In the spatially flat gauge, the primordial curvature perturbation is defined as
\begin{align}
\zeta=-\frac{H}{\dot{\rho}_{\mathrm{tot}} }  \delta \rho, 
\end{align}
where $\delta\rho$ is the unperturbed energy density perturbation and $\rho_{\mathrm{tot}}$ denotes the total energy density, 
including contributions from the background inflaton $\phi$ and the backreaction of $\chi$. Since the $\chi$ field lacks a homogeneous background component, 
it does not contribute directly to $\delta\rho$.
Accordingly, the power spectrum of primordial curvature perturbation in the super-horizon regime, including rescattering effects, reads
\begin{align}
 P_{\zeta}^{(1,2)}=\frac{H^2 p^3}{2\pi^2\left\langle\dot{\phi}\right\rangle^2}\left|\delta\phi_p\right|_{(1,2)}^2,      
\end{align}
where $\left\langle\dot{\phi}\right\rangle$ denotes the inflaton velocity modified by backreaction. 
The leading-order backreaction correction to the background inflaton $\phi$  arises from the tadpole diagram and is expressed as
\begin{align}
\label{tadpole}
\left\langle\delta \phi_{\boldsymbol{p}}(\tau)\right\rangle&=\left\langle i\int^{\tau}_{\tau_*} d \tau^{\prime} H_I^{(3)}(\tau^\prime)\delta \phi_{\boldsymbol{p}}(\tau)-\int^{\tau}_{\tau_*} d \tau^{\prime}\delta \phi_{\boldsymbol{p}}(\tau)H_I^{(3)}(\tau^{\prime})\right\rangle \nonumber\\&=-\delta^{(3)}\left(\boldsymbol{p}\right)g\int^{\tau}_{\tau_*} d \tau^{\prime} a^2(\tau^{\prime}) \left[\phi(\tau^{\prime})\!-\!\phi_0\right]G_p(\tau,\tau^{\prime})\int dk^3f_k(\tau^{\prime},\tau^{\prime}).
\end{align}

\begin{figure}
\centering
\includegraphics[width=0.25\textheight]{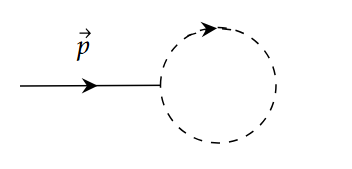}
\caption{The Feynman diagram of the tadpole~\eqref{tadpole}.  }
\label{fig2}
\end{figure}

\begin{figure}
\centering
\includegraphics[width=0.5\textheight]{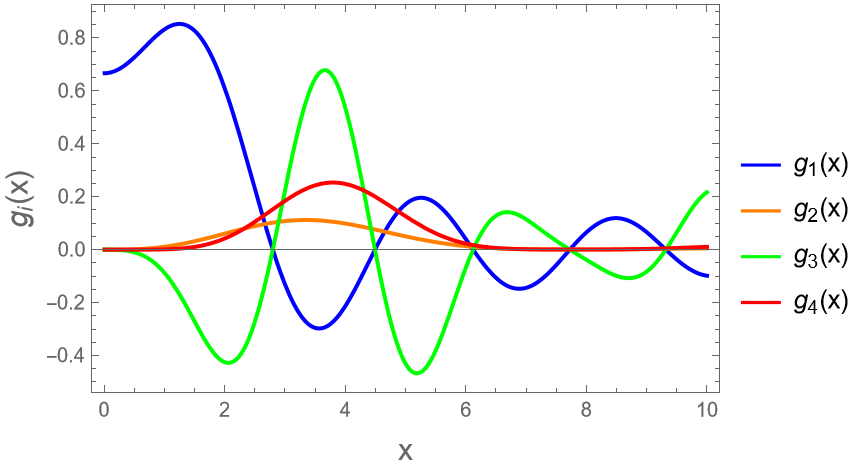}
\caption{Functional forms of $g_1(x)$ - $g_4(x)$, representing the shapes of the spectra.  }
\label{fig3}
\end{figure}

The Feynman diagram for the tadpole is shown in Fig~\ref{fig2}. 
To evaluate this, we neglect the oscillatory contributions within the function $f_k(\tau^{\prime},\tau^{\prime})$ when performing the integral. 
Utilizing Eqs.\eqref{omega_3} and\eqref{beta}, the tadpole correction becomes
\begin{align}
\left\langle\delta \phi(\tau)\right\rangle&=\int \frac{dp^3}{(2\pi)^3} e^{i\boldsymbol{p}\cdot \boldsymbol{x}}\left\langle\delta \phi_{\boldsymbol{p}}(\tau)\right\rangle \nonumber \\&  
=-\frac{g}{16 \pi^3 } \frac{(-g \dot{\phi}_0)^{\frac{3}{2}}}{H^3 \tau_0^3}e^{\frac{2 \pi(1+6 \xi) H^2}{-g \dot{\phi}_0}} \int^{\tau}_{\tau_*} d \tau_1 a(\tau_1)\frac{H^2\left(\tau^3-\tau_1^{3}\right)}{6},
\end{align}
Taking the time derivative of this equation, we obtain
\begin{equation}
\begin{split}
\left\langle\delta \dot{\phi}\right\rangle&=\frac{g(-g \dot{\phi}_0)^{\frac{3}{2}}}{16 \pi^3 H }\exp\left[-\frac{2 \pi(1+6 \xi) H^2}{g \dot{\phi}_0}\right]\left( \frac{a_0}{a}\right)^3\ln{\frac{a}{a_i}}  \\
&\approx \frac{g(-g \dot{\phi}_0)^{\frac{3}{2}}}{16 \pi^3  }\exp\left[-\frac{2 \pi(1+6 \xi) H^2}{g \dot{\phi}_0}\right](t-t_i)e^{-3H(t-t_i)} ,
\end{split}
\end{equation}
After particle production occurs for times $t>t_0+\Delta t/2$, 
the expectation value $\langle\delta \dot{\phi}\rangle$ decays as $ t e^{-3Ht}$,
becoming negligible after several Hubble times. This behavior is consistent with numerical results reported in~\cite{Yu:2023ity}. 
Consequently, backreaction effects in the superhorizon regime can be safely neglected, 
allowing us to approximate $\langle \dot{\phi}\rangle\approx\dot{\phi}$. 
We then compute the ratio of the power spectrum including rescattering effects to the unperturbed power spectrum in the superhorizon regime as
\begin{equation}
\delta^{(1,2)}_{\delta \phi}=\frac{2 p^3}{H^2}\left|\delta\phi_p\right|_{(1,2)}^2,  
\end{equation}
To derive precise expressions, we again employ the approximation \eqref{omega_3}. 
The integrand term $f_p(\tau^\prime,\tau^{\prime\prime})f_{|\boldsymbol{k}-\boldsymbol{p}|}(\tau^\prime,\tau^{\prime\prime})$ in Eq. \eqref{u2} complicates the integration. 
However, the dominant contribution to the momentum integrals originates from internal momenta satisfying $k^2 \sim 12\xi H^2 a^2_0$. 
Given that the external momentum $p$ peaks at $p_0 = H a_0$, 
this implies $k \sim \sqrt{12\xi} p_0$. 
Numerical studies \cite{Yu:2023ity} indicate that particle production is maximized for $\xi \sim \mathcal{O}(1)$, 
implying that $|\boldsymbol{k}-\boldsymbol{p}| \sim \mathcal{O}(k)$. 
Thus, we approximate $f_{|\boldsymbol{k}-\boldsymbol{p}|}(\tau^\prime,\tau^{\prime\prime}) \approx f_k(\tau^\prime,\tau^{\prime\prime})$. By neglecting rapidly oscillating and subdominant contributions, we get the final results,
\begin{align}
\delta^{(1)}_{\delta \phi}=\frac{g^{2} (-g\dot{ \phi}_{0})^{\frac{1}{2}}}{8\pi^3 H}e^{ \frac{2 \pi \left( 1+6 \xi \right) H^{2}}{-g \dot{ \phi}_{0}}}g_1(x)\ln{\gamma},    
\end{align}
\begin{align}
\delta^{(2)}_{\delta \phi}=\frac{g^2(-g\dot{\phi}_0)^{\frac{3}{2}} }{4 \pi^3 H^3}e^{ \frac{2 \pi \left( 1+6 \xi \right) H^{2}}{-g \dot{ \phi}_{0}}}\left(1+\frac{1}{\sqrt{2}}e^{ \frac{2 \pi \left( 1+6 \xi \right) H^{2}}{-g \dot{ \phi}_{0}}}\right)g_2(x),    
\end{align}
where
\begin{align}
g_1(x)=\frac{-2x\cos{2x}+\left(1-x^2\right)\sin{2x}}{x^3} ,\qquad g_2(x)=\frac{\left(\sin{x}-\int^x_0 \frac{dx\sin{x}}{x}\right)^2}{x^3} ,
\end{align}
and $x \equiv p/p_0$. 
The functional forms of $g_1(x)$ and $g_2(x)$ are identical to those obtained in the minimal coupling case \cite{Pearce:2017bdc} and are illustrated in Fig.~\ref{fig3}. 
Notably, $g_1(x)$ features an oscillatory structure stemming from the quartic interaction term $H_I^{(4)}$, 
while $g_2(x)$ exhibits a characteristic bump.
Their respective peak values occur at $g_1(1.25) \approx 0.8531$ and $g_2(3.34) \approx 0.112$, 
consistent with expectations from particle production phenomena.
Compared to the minimal case, the amplitude of the power spectrum is significantly enhanced.
This enhancement can be quantitatively estimated as follows. 
In our study, particle production occurs predominantly on small scales, leaving the power spectrum on CMB scales essentially unaltered. 
Therefore, the Hubble parameter $H$ can be constrained by the observed amplitude of the power spectrum at the pivot scale $k = 0.05 \mathrm{Mpc}^{-1}$,
together with the tensor-to-scalar ratio $r$ measured at the same scale, 
resulting in an upper bound $H \leq 2 \times 10^{-5}$. 
Despite potentially strong backreaction effects at the particle production time $t_0$, 
the Universe remains in its inflationary phase. 
Under the slow-roll approximation, we can thus express the background inflaton velocity $\dot{\phi}_0$ and the slope of the potential $\partial_\phi V(\phi_0)$ approximately in terms of the slow-roll parameter $\epsilon_0$, as $\dot{\phi}_0\simeq\sqrt{2\epsilon_0}$ and $\partial_\phi V(\phi_0)\simeq 3H^2\sqrt{2\epsilon_0}$. 
Using these approximations, the peak values of the power spectrum ratios $\delta^{(1,2)}_{\delta \phi}$ can be estimated as
\begin{equation}
 \delta^{(1)}_{\delta \phi}\approx-5.12*\ln{\gamma} \geq 3.55,  
\end{equation}
\begin{equation}
\delta^{(2)}_{\delta \phi}\approx 54.28\frac{\pi^3\epsilon_0^{\frac{1}{4}}}{g^{\frac{3}{2}}H^{\frac{1}{2}}}.
\end{equation}
To estimate the magnitude of $\delta^{(2)}_{\delta\phi}$,
we consider representative values appropriate to small-scale modes during inflation, $H = 10^{-7}$ and $\epsilon_0 = 10^{-4}$.
Applying the particle production duration constraint from Eq.~\eqref{time}, 
the maximum $\delta^{(2)}_{\delta\phi}$ is found to be of order $1.63 \times 10^{12}$ at a coupling $g = 0.00204$. 
This large value indicates that there exists a viable parameter space where the power spectrum can be sufficiently enhanced to induce PBHs formation and generate accompany SIGWs.
Choosing a larger coupling $g=0.1$ results in $\gamma\approx0.01267$, consistent with the duration constraint, 
and yields $\delta^{(2)}_{\delta \phi}\approx 1.68\times 10^7$. 
Under the same parameters, the contribution from
$\delta^{(1)}_{\delta\phi}$ is much smaller, approximately $15.6$, and thus negligible compared to $\delta^{(2)}_{\delta\phi}$. 
Consequently, when combined with the unperturbed small-scale power spectrum $P_{\zeta}^{(0)} \sim 10^{-9}$,
the total power spectrum including rescattering effects can be enhanced up to $P_{\zeta}^{(2)} \sim 10^{-2}$,
sufficiently large to support PBH formation and generate significant SIGWs.

\subsection{Non-Gaussian}
\label{sec:23}
\begin{figure}
\centering
\includegraphics[width=0.25\textheight]{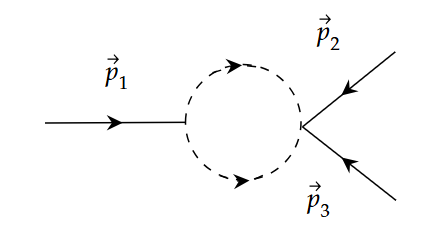}
\includegraphics[width=0.2\textheight]{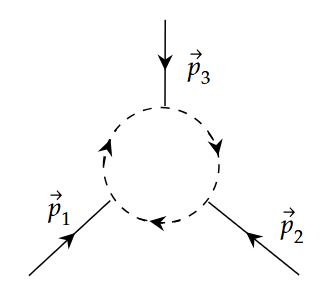}
\caption{Feynman diagrams illustrating the non-Gaussian contributions from the rescattering of particle $\chi$. The left and right panels correspond to Eq.~\eqref{loop4} and Eq.~\eqref{loop5}, respectively.  }
\label{fig4}
\end{figure}

In this subsection, we investigate the non-Gaussianity of curvature perturbations induced by rescattering. 
Our analytical approach parallels the methodology employed for power spectrum corrections in the previous section, 
with our analysis focused on the equilateral bispectrum configuration. 
Substituting $Q=\delta \phi_{\boldsymbol{p}_1} \delta \phi_{\boldsymbol{p}_2}\delta \phi_{\boldsymbol{p}_3}$ into the in-in formula~\eqref{inin}, the resulting bispectrum from rescattering processes is expressed as
\begin{equation}
\label{loop4}
\begin{split}
&\left\langle\delta \phi_{\boldsymbol{p}_1}(\tau) \delta \phi_{\boldsymbol{p}_2}(\tau)\delta \phi_{\boldsymbol{p}_3}(\tau)\right\rangle_1\\
&=\left\langle\int^\tau_{\tau_*} d \tau^{\prime} H_I^{(3)}(\tau^{\prime}) \delta \phi_{\boldsymbol{p}_1}(\tau) \delta \phi_{\boldsymbol{p}_2}(\tau)\delta \phi_{\boldsymbol{p}_3}(\tau) \int^\tau_{\tau_*} d \tau^{\prime \prime} H_I^{(4)}(\tau^{\prime \prime}) \right\rangle
 \\
 &\quad+ \left\langle\int^\tau_{\tau_*} d \tau^{\prime} H_I^{(4)}(\tau^{\prime}) \delta \phi_{\boldsymbol{p}_1}(\tau) \delta \phi_{\boldsymbol{p}_2}(\tau)\delta \phi_{\boldsymbol{p}_3}(\tau) \int^\tau_{\tau_*} d \tau^{\prime \prime} H_I^{(3)}(\tau^{\prime \prime}) \right\rangle
 \\
 & \quad-   \left\langle\int^\tau_{\tau_*} d \tau^{\prime} \int^{\tau^{\prime}}_{\tau_*} d \tau^{\prime \prime}\left[H^{(3)}(\tau^{\prime\prime}) H^{(4)}(\tau^{\prime})+H^{(4)}(\tau^{\prime\prime}) H^{(3)}(\tau^{\prime})\right]\delta \phi_{\boldsymbol{p}_1}(\tau) \delta \phi_{\boldsymbol{p}_2}(\tau)\delta \phi_{\boldsymbol{p}_3}(\tau)\right\rangle  \\
 & \quad - \left\langle\int^\tau_{\tau_*} d \tau^{\prime} \int^{\tau^{\prime}}_{\tau_*} d \tau^{\prime \prime} \delta \phi_{\boldsymbol{p}_1}(\tau) \delta \phi_{\boldsymbol{p}_2}(\tau)\delta \phi_{\boldsymbol{p}_3}(\tau)\left[ H^{(3)}(\tau^{\prime}) H^{(4)}(\tau^{\prime \prime})+ H^{(4)}(\tau^{\prime}) H^{(3)}(\tau^{\prime \prime})\right]\right\rangle\\
&=(2\pi)^3\delta^{(3)}\left(\boldsymbol{p}_1+\boldsymbol{p}_2+\boldsymbol{p}_3\right)B^{(1)}(p_1,p_2,p_3),
\end{split}
\end{equation}

\begin{equation}
\label{loop5}
\begin{split}
&\left\langle\delta \phi_{\boldsymbol{p}_1}(\tau) \delta \phi_{\boldsymbol{p}_2}(\tau)\delta \phi_{\boldsymbol{p}_3}(\tau)\right\rangle_2\nonumber\\&=-i \left\langle\int^\tau_{\tau_*} d \tau^{\prime} \int^{\tau^{\prime}}_{\tau_*} d \tau^{\prime \prime} \int^{\tau^{\prime\prime}}_{\tau_*} d \tau^{\prime \prime\prime}\left[ H^{(3)}(\tau^{\prime\prime\prime}),\left[ H^{(3)}(\tau^{\prime\prime}),\left[ H^{(3)}(\tau^{\prime}),\delta \phi_{\boldsymbol{p}_1}(\tau) \delta \phi_{\boldsymbol{p}_2}(\tau)\delta \phi_{\boldsymbol{p}_3}(\tau)\right]\right] \right]\right\rangle\\
&=(2\pi)^3\delta^{(3)}\left(\boldsymbol{p}_1+\boldsymbol{p}_2+\boldsymbol{p}_3\right)B^{(2)}(p_1,p_2,p_3) 
\end{split}
\end{equation}
The corresponding diagrammatic representations are shown in Fig. \ref{fig4}. 
Utilizing these Feynman diagrams together with the propagators and Green’s functions defined in Eqs. \eqref{Pro_phi}–\eqref{Green_chi}, 
we derive the detailed bispectrum expression,
\begin{align}
 B^{(1)}(p_1, p_2, p_3)=&-24 g^4 \int^\tau_{\tau_*} d \tau_1 d \tau_2 a^2(\tau_1)a^2(\tau_2)\left(\phi(\tau_1)\!-\!\phi_0\right)\left(\phi(\tau_2)\!-\!\phi_0\right)Im\left[\delta\phi_{p_1}(\tau_1)\delta\phi^\ast_{p_2}(\tau)\right] \nonumber\\&\times Im\left[\delta\phi_{p_2}(\tau)\delta\phi^\ast_{p_2}(\tau_2)\delta\phi_{p_3}(\tau)\delta\phi^\ast_{p_3}(\tau_2)\right]\int \frac{dk^3}{(2\pi)^3}f_k(\tau_1, \tau_2)f_{|\boldsymbol{k}-\boldsymbol{p}_1|}(\tau_2, \tau_1),
\end{align}
\begin{align}
 B^{(2)}(p_1, p_2, p_3)=&-8 g^6 \int^\tau_{\tau_*} d \tau_1 d \tau_2 d \tau_3a^2(\tau_1)a^2(\tau_2)a^2(\tau_3)\left(\phi(\tau_1)\!-\!\phi_0\right)\left(\phi(\tau_2)\!-\!\phi_0\right)\left(\phi(\tau_3)\!-\!\phi_0\right) \nonumber\\&\times G_{p_1}(\tau_1,\tau)G_{p_2}(\tau_2,\tau)G_{p_3}(\tau_3,\tau)\int \frac{dk^3}{(2\pi)^3}f_k(\tau_1, \tau_2)f_{|\boldsymbol{k}+\boldsymbol{p}_2|}(\tau_2, \tau_3)f_{|\boldsymbol{k}-\boldsymbol{p}_1|}(\tau_1, \tau_3).
\end{align}
Given the consistent validity of the approximation $f_{|\boldsymbol{k}\pm\boldsymbol{p}|}(\tau_1,\tau_2) \approx f_k(\tau_1,\tau_2)$ and neglecting rapidly oscillating terms, the equilateral bispectrum (with $p_1 = p_2 = p_3 = p$) reduces to
\begin{align}
B^{(1)}(p, p, p)=-\frac{3g^{3}(-g\dot{\phi}_0)^{\frac{1}{2}}H^2}{16\pi^3 p^6}e^{ \frac{2 \pi \left( 1+6 \xi \right) H^{2}}{-g \dot{ \phi}_{0}}}\left(1+\frac{1}{\sqrt{2}}e^{ \frac{2 \pi \left( 1+6 \xi \right) H^{2}}{-g \dot{ \phi}_{0}}}\right)g_3(x)\ln{\gamma},    
\end{align}
\begin{align}
B^{(2)}(p, p, p)=\frac{g^3(-g\dot{\phi}_0)^{\frac{3}{2}} }{ \pi^3 p^6}e^{ \frac{4 \pi \left( 1+6 \xi \right) H^{2}}{-g \dot{ \phi}_{0}}}\left(\frac{3}{8\sqrt{2}}+\frac{1}{3\sqrt{3}}e^{ \frac{2 \pi \left( 1+6 \xi \right) H^{2}}{-g \dot{ \phi}_{0}}}\right)g_4(x),    
\end{align}
where
\begin{align}
g_3(x)=\frac{(2x\cos{2x}-\left(1-x^2\right)\sin{2x})\left(\sin{x}-\int^x_0 \frac{dx\sin{x}}{x}\right)}{x^3} ,\quad g_4(x)=\frac{\left(\sin{x}-\int^x_0 \frac{dx\sin{x}}{x}\right)^3}{x^3} .
\end{align}
The functional forms of $g_3(x)$ and $g_4(x)$ are plotted in Fig. \ref{fig3}. 

Similar to $g_1(x)$, $g_3(x)$ exhibits an oscillatory structure stemming from the quartic interaction term $H^{(4)}_I$, 
peaking at $g_3(3.662) = 0.6782$. 
Meanwhile, $g_4(x)$ features a narrow bump, attaining a maximum of $g_4(3.796) = 0.2536$. 
To connect these bispectrum contributions to observations, 
we introduce the standard nonlinearity parameter $f_\text{NL}$ \cite{Chen:2006nt},
\begin{align}
B(p_1,p_2,p_3)=\frac{24}{5}\pi^4 f_{\mathrm{NL}}(p_1,p_2,p_3)\left(P_{\zeta,\,p_1}^{(0)}P_{\zeta,\,p_2}^{(0)}+P_{\zeta,\,p_2}^{(0)}P_{\zeta,\,p_3}^{(0)}+P_{\zeta,\,p_1}^{(0)}P_{\zeta,\,p_3}^{(0)}\right)\frac{p_1^3+p_2^3+p^3_3}{p_1^3p_2^3p_3^3},    
\end{align}
Employing the assumptions and methods established in Section \ref{sec:22}, 
the nonlinearity parameters for the two bispectrum contributions can be expressed in terms of slow-roll parameters as
\begin{align}
f_{\mathrm{NL}}^{(1)}=-\left(\frac{5gH^3}{24\pi^4}+\frac{5\sqrt{2}H^\frac{7}{2}}{\pi g^{\frac{3}{2}}(2\epsilon_0)^{\frac{1}{4}}}\right)\frac{g_3(x)\ln{\gamma}}{|P^{(0)}_\zeta|^2} \approx 1.28 \frac{H^\frac{7}{2}}{ g^{\frac{3}{2}}\epsilon_0^{\frac{1}{4}}}\frac{\ln{\gamma}}{|P^{(0)}_\zeta|^2}
\end{align}
\begin{align}
f_{\text{NL}}^{(2)}=(\frac{20H^{5/2}(2\epsilon_0)^{1/4}}{\sqrt{2}\pi g^{1/2}}+\frac{2560\pi^2H^3}{3\sqrt{3}g^3})\frac{g_4(x)}{|P^{(0)}_\zeta|^2}\approx   1233\frac{H^3}{g^3|P^{(0)}_\zeta|^2} ,
\end{align}
Adopting the parameter values consistent with those producing an enhanced power spectrum of order $\mathcal{O}(10^{-2})$, we obtain $f_{\text{NL}}^{(1)}=5.6\times 10^{-4}$ and $f_{\text{NL}}^{(2)}=1233.12$.
This large value of $f_{\text{NL}}^{(2)}$ indicates significant non-Gaussianity arising from the rescattering processes, 
which could have important phenomenological consequences for PBH formation and SIGWs.

\section{Conclusion and Discussion}
\label{sec:3}

In this work, we have investigated rescattering processes resulting from sudden particle production with non-minimal coupling during inflation. 
The tachyonic instability inherent in this scenario produces particle numbers vastly exceeding those arising in minimal coupling cases, 
leading to notably strong rescattering effects. By incorporating the backreaction of produced particles on the background dynamics, 
we computed curvature perturbation power spectra including one-loop rescattering corrections. 
Our results demonstrate that the power spectrum $P_{\zeta}$ can be enhanced up to $\mathcal{O}(10^{-2})$ in suitable parameter regimes near the scales of particle production. 
Such an amplitude is sufficient to generate both PBHs and SIGWs.
For the same parameter sets that enhance the power spectra, 
we also quantified the non-Gaussianity via the nonlinearity parameter $f_{\text{NL}}$, 
finding enhancements reaching $\mathcal{O}(10^{3})$ at particle production scales. 
These large values of $f_{\text{NL}}$ are characteristic of the model and may further facilitate PBH formation.

Several limitations of our study merit attention: (1) The analytical approximations used may lead to overestimation of the effects; 
(2) The shape functions $g_1(x)$ and $g_2(x)$ do not include slow-roll corrections, 
limiting their applicability for spectral index computations at CMB scales; 
(3) Backreaction has been treated only to lowest order, 
thus providing upper bounds on the parameter$\xi$; 
and (4) inclusion of the interaction term \eqref{Intaction_final} could modify $\langle\chi^2\rangle$, and consequently affect the final power spectra. 

Future work will employ lattice simulations~\cite{Caravano:2024moy,Caravano:2025diq} to overcome these limitations, 
enabling precise determination of spectral indices, PBH mass distributions, and SIGW signals.

\begin{acknowledgments}
We thank Yigan Zhu for useful discusstions. This work is supported by the National Natural Science Foundation of China under Grant No. 12447176.
\end{acknowledgments}

\appendix

\section{\label{A}The dominant interactions}

This appendix derives the dominant interaction terms relevant during slow-roll inflation.
Expanding the action \eqref{action} to cubic and quartic orders in perturbations yields the interaction Hamiltonian density
\begin{align}
\label{Interaction}
\mathcal{H}_I=a^3 & {\left[-g^2\left(\phi-\phi_0\right) \delta \phi \chi^2-\frac{g^2}{2} \delta \phi^2 \chi^2-\frac{1}{2 a^2} \Psi \dot{\chi}^2-\frac{1}{2 a^2} \Psi \partial_i \chi \partial_i \chi-\frac{g^2}{2} \Psi\left(\phi-\phi_0\right)^2 \chi^2\right.}\nonumber \\
& \left.-\frac{1}{a} \dot{\chi} \partial_i \chi \partial_i B+\frac{\xi}{2}R^{(1)}\chi^2+\frac{\xi}{2}R^{(2)}\chi^2\right]\nonumber,
\end{align}
where the linear and quadratic parts of the Ricci scalar perturbation are
\begin{align}
R^{(1)}=-24 H^2 \Psi-12 \Psi \dot{H}-\frac{6 H\partial_i \partial_i B}{a}-\frac{2\partial_{i} \partial_{i} \dot{B}}{a}-6 H \dot{\Psi}-\frac{2\partial_i \partial_i \Psi}{a^2},    
\end{align}
\begin{align}
R^{(2)}=&4 H^2\left(\partial_i B\right)^2+2 \dot{H}\left(\partial_i B\right)^2-6 H\partial_i B\partial_i \dot{B}+\frac{12 H \Psi\partial_i \partial_i B}{a}-\frac{\left(\partial_i \partial_j B\right)^2}{a^2}+\frac{\left(\partial_i \partial_i B\right)^2}{a^2}\nonumber\\&+\frac{4 \Psi\partial_i \partial_i \dot{B}}{a}+24 H \Psi \dot{\Psi}+\frac{2\partial_i \partial_i B \dot{\Psi}}{a}+\frac{6 H\partial_i B\partial_i \Psi}{a}+\frac{2\left(\partial_i \Psi\right)^2}{a^2}+\frac{4 \Psi\partial_i \partial_i \Psi}{a^2}.
\end{align}
Following the approach of Ref.~\cite{Pearce:2017bdc}, the constraint equations governing first-order metric perturbations are
\begin{align}
\label{constraint1}
 2HB+\dot{B}+\frac{\Psi}{a}=0 ,  
\end{align}
\begin{align}
\Psi=\frac{\dot{\phi}\delta \phi}{H},
\end{align}
\begin{align}
H\dot{\Psi}=-\left(3+\frac{\dot{H}}{H^2}\right) \frac{\dot{\phi} H}{2} \delta \phi+\frac{1}{2}\left(\dot{\phi} \delta \dot{\phi}-\frac{\partial V}{\partial \phi} \delta \phi\right),  
\end{align}
\begin{align}
-\frac{2 H}{a} \partial_i \partial_i B=\dot{\phi} \delta \dot{\phi}+\frac{\dot{\phi} V}{H} \delta \phi+\frac{\partial V}{\partial \phi} \delta \phi.
\end{align}
Using the slow-roll conditions $\dot{\phi}=- \sqrt{2 \epsilon} H$ and $\ddot{\phi}=-\dot{\phi}H(\epsilon+\eta)$, 
these equations can be further simplified as
\begin{align}
\label{constraint2}
\Psi=-\frac{\sqrt{2\epsilon}}{2}\delta\phi,
\end{align}
\begin{align}
\label{constraint3}
\dot{\Psi}=-\frac{\sqrt{2\epsilon}}{2}\left[\delta\dot{\phi}+H(\eta-2\epsilon)\delta\phi\right],
\end{align}
\begin{align}
\label{constraint4}
-\frac{1}{a} \partial_i \partial_i B=-\frac{\sqrt{2\epsilon}}{2}\left[\delta \dot{\phi}+(\eta-2\epsilon) H\delta \phi\right].
\end{align}
Substituting these relations back into the interaction Hamiltonian density \eqref{Interaction}, 
one finds that only four terms dominate,
\begin{align}
 \mathcal{H}_I\approx a^3\left[-g^2\left(\phi-\phi_0\right) \delta \phi \chi^2-\frac{g^2}{2} \delta \phi^2 \chi^2+\frac{\xi}{2}R^{(1)}\chi^2+\frac{\xi}{2}R^{(2)}\chi^2\right], \nonumber
\end{align}
The last two terms involving $R^{(1)}$ and $R^{(2)}$ deserve careful consideration. 
First, using Eq. \eqref{constraint1} to eliminate $\dot{B}$,
and neglecting terms proportional to $\dot{H}$ under the slow-roll approximation, we simplify them as 
\begin{align}
R^{(1)}\approx-24 H^2 \Psi-\frac{2 H\partial_i \partial_i B}{a}-6 H \dot{\Psi}, \nonumber
\end{align}
\begin{align}
R^{(2)}\approx&16 H^2\left(\partial_i B\right)^2+\frac{12 H\partial_i B\partial_i \Psi}{a}+\frac{2\left(\partial_i \Psi\right)^2}{a^2}+\frac{8 H \Psi\partial_i \partial_i B}{a}\nonumber\\&-\frac{\left(\partial_i \partial_j B\right)^2}{a^2}+\frac{\left(\partial_i \partial_i B\right)^2}{a^2}+24 H \Psi \dot{\Psi}+\frac{2\partial_i \partial_i B \dot{\Psi}}{a}, \nonumber
\end{align}
Substituting Eqs. \eqref{constraint2}–\eqref{constraint4} into these expressions and discarding higher-order slow-roll corrections, 
the dominant contribution in  $R^{(1)}$ reduces to $-24 H^2 \Psi = 12\sqrt{2\epsilon} H^2 \delta\phi$.
The expression for $R^{(2)}$ is more involved, but its leading terms are approximately
\begin{align}
R^{(2)}\approx&16 H^2\left(\partial_i B\right)^2+\frac{12 H\partial_i B\partial_i \Psi}{a}+\frac{2\left(\partial_i \Psi\right)^2}{a^2}+\frac{8 H \Psi\partial_i \partial_i B}{a}\nonumber+24 H \Psi \dot{\Psi}, \nonumber
\end{align}
The first three terms are of comparable magnitude, 
estimated as  $\epsilon k^2 a^{-2} \delta\phi^2$. 
Similarly, the last two terms scale as
$\epsilon \left( \delta\dot{\phi} \delta\phi + (\eta - 2\epsilon) H \delta\phi^2 \right)$. 
The validity condition for sudden particle production, inequality \eqref{time}, implies the hierarchy
$g^2 > (6\xi+1) H^2 / \epsilon_0 > \sqrt{\epsilon} \, \xi H^2$,
which ensures the contributions from the terms $\xi R^{(1)} \delta\phi^2 \chi^2$ and $\xi R^{(2)} \delta\phi^2 \chi^2$ are negligible compared to the leading interactions.

\bibliographystyle{apsrev4-2f}
\bibliography{references} 

\begin{thebibliography}{45}%
\makeatletter
\providecommand \@ifxundefined [1]{%
 \@ifx{#1\undefined}
}%
\providecommand \@ifnum [1]{%
 \ifnum #1\expandafter \@firstoftwo
 \else \expandafter \@secondoftwo
 \fi
}%
\providecommand \@ifx [1]{%
 \ifx #1\expandafter \@firstoftwo
 \else \expandafter \@secondoftwo
 \fi
}%
\providecommand \natexlab [1]{#1}%
\providecommand \enquote  [1]{``#1''}%
\providecommand \bibnamefont  [1]{#1}%
\providecommand \bibfnamefont [1]{#1}%
\providecommand \citenamefont [1]{#1}%
\providecommand \href@noop [0]{\@secondoftwo}%
\providecommand \href [0]{\begingroup \@sanitize@url \@href}%
\providecommand \@href[1]{\@@startlink{#1}\@@href}%
\providecommand \@@href[1]{\endgroup#1\@@endlink}%
\providecommand \@sanitize@url [0]{\catcode `\\12\catcode `\$12\catcode `\&12\catcode `\#12\catcode `\^12\catcode `\_12\catcode `\%12\relax}%
\providecommand \@@startlink[1]{}%
\providecommand \@@endlink[0]{}%
\providecommand \url  [0]{\begingroup\@sanitize@url \@url }%
\providecommand \@url [1]{\endgroup\@href {#1}{\urlprefix }}%
\providecommand \urlprefix  [0]{URL }%
\providecommand \Eprint [0]{\href }%
\providecommand \doibase [0]{https://doi.org/}%
\providecommand \selectlanguage [0]{\@gobble}%
\providecommand \bibinfo  [0]{\@secondoftwo}%
\providecommand \bibfield  [0]{\@secondoftwo}%
\providecommand \translation [1]{[#1]}%
\providecommand \BibitemOpen [0]{}%
\providecommand \bibitemStop [0]{}%
\providecommand \bibitemNoStop [0]{.\EOS\space}%
\providecommand \EOS [0]{\spacefactor3000\relax}%
\providecommand \BibitemShut  [1]{\csname bibitem#1\endcsname}%
\let\auto@bib@innerbib\@empty
\bibitem [{\citenamefont {Guth}(1981)}]{Guth:1980zm}%
  \BibitemOpen
  \bibfield  {author} {\bibinfo {author} {\bibfnamefont {A.~H.}\ \bibnamefont {Guth}},\ }\bibinfo {title} {{The Inflationary Universe: A Possible Solution to the Horizon and Flatness Problems}},\ \href {https://doi.org/10.1103/PhysRevD.23.347} {\bibfield  {journal} {Phys. Rev. D\ }\textbf {\bibinfo {volume} {23}},\ \bibinfo {pages} {347} (\bibinfo {year} {1981})}\BibitemShut {NoStop}%
\bibitem [{\citenamefont {Starobinsky}(1980)}]{Starobinsky:1980te}%
  \BibitemOpen
  \bibfield  {author} {\bibinfo {author} {\bibfnamefont {A.~A.}\ \bibnamefont {Starobinsky}},\ }\bibinfo {title} {{A New Type of Isotropic Cosmological Models Without Singularity}},\ \href {https://doi.org/10.1016/0370-2693(80)90670-X} {\bibfield  {journal} {Phys. Lett. B\ }\textbf {\bibinfo {volume} {91}},\ \bibinfo {pages} {99} (\bibinfo {year} {1980})}\BibitemShut {NoStop}%
\bibitem [{\citenamefont {Sato}(1981)}]{Sato:1980yn}%
  \BibitemOpen
  \bibfield  {author} {\bibinfo {author} {\bibfnamefont {K.}~\bibnamefont {Sato}},\ }\bibinfo {title} {{First Order Phase Transition of a Vacuum and Expansion of the Universe}},\ \href@noop {} {\bibfield  {journal} {Mon. Not. R. Astron. Soc.\ }\textbf {\bibinfo {volume} {195}},\ \bibinfo {pages} {467} (\bibinfo {year} {1981})}\BibitemShut {NoStop}%
\bibitem [{\citenamefont {Linde}(1982)}]{Linde:1981mu}%
  \BibitemOpen
  \bibfield  {author} {\bibinfo {author} {\bibfnamefont {A.~D.}\ \bibnamefont {Linde}},\ }\bibinfo {title} {{A New Inflationary Universe Scenario: A Possible Solution of the Horizon, Flatness, Homogeneity, Isotropy and Primordial Monopole Problems}},\ \href {https://doi.org/10.1016/0370-2693(82)91219-9} {\bibfield  {journal} {Phys. Lett. B\ }\textbf {\bibinfo {volume} {108}},\ \bibinfo {pages} {389} (\bibinfo {year} {1982})}\BibitemShut {NoStop}%
\bibitem [{\citenamefont {Mukhanov}\ and\ \citenamefont {Chibisov}(1981)}]{Mukhanov:1981xt}%
  \BibitemOpen
  \bibfield  {author} {\bibinfo {author} {\bibfnamefont {V.~F.}\ \bibnamefont {Mukhanov}}\ and\ \bibinfo {author} {\bibfnamefont {G.~V.}\ \bibnamefont {Chibisov}},\ }\bibinfo {title} {{Quantum Fluctuations and a Nonsingular Universe}},\ \href@noop {} {\bibfield  {journal} {JETP Lett.\ }\textbf {\bibinfo {volume} {33}},\ \bibinfo {pages} {532} (\bibinfo {year} {1981})}\BibitemShut {NoStop}%
\bibitem [{\citenamefont {Sasaki}(1983)}]{Sasaki:1983kd}%
  \BibitemOpen
  \bibfield  {author} {\bibinfo {author} {\bibfnamefont {M.}~\bibnamefont {Sasaki}},\ }\bibinfo {title} {{Gauge Invariant Scalar Perturbations in the New Inflationary Universe}},\ \href {https://doi.org/10.1143/PTP.70.394} {\bibfield  {journal} {Prog. Theor. Phys.\ }\textbf {\bibinfo {volume} {70}},\ \bibinfo {pages} {394} (\bibinfo {year} {1983})}\BibitemShut {NoStop}%
\bibitem [{\citenamefont {Kodama}\ and\ \citenamefont {Sasaki}(1984)}]{Kodama:1984ziu}%
  \BibitemOpen
  \bibfield  {author} {\bibinfo {author} {\bibfnamefont {H.}~\bibnamefont {Kodama}}\ and\ \bibinfo {author} {\bibfnamefont {M.}~\bibnamefont {Sasaki}},\ }\bibinfo {title} {{Cosmological Perturbation Theory}},\ \href {https://doi.org/10.1143/PTPS.78.1} {\bibfield  {journal} {Prog. Theor. Phys. Suppl.\ }\textbf {\bibinfo {volume} {78}},\ \bibinfo {pages} {1} (\bibinfo {year} {1984})}\BibitemShut {NoStop}%
\bibitem [{\citenamefont {Akrami}\ {\it et~al.}(2020)\citenamefont {Akrami} {\it et~al.}}]{Planck:2018jri}%
  \BibitemOpen
  \bibfield  {author} {\bibinfo {author} {\bibfnamefont {Y.}~\bibnamefont {Akrami}} {\it et~al.} (\bibinfo {collaboration} {Planck}),\ }\bibinfo {title} {{Planck 2018 results. X. Constraints on inflation}},\ \href {https://doi.org/10.1051/0004-6361/201833887} {\bibfield  {journal} {Astron. Astrophys.\ }\textbf {\bibinfo {volume} {641}},\ \bibinfo {pages} {A10} (\bibinfo {year} {2020})}\BibitemShut {NoStop}%
\bibitem [{\citenamefont {Aghanim}\ {\it et~al.}(2020)\citenamefont {Aghanim} {\it et~al.}}]{Planck:2018vyg}%
  \BibitemOpen
  \bibfield  {author} {\bibinfo {author} {\bibfnamefont {N.}~\bibnamefont {Aghanim}} {\it et~al.} (\bibinfo {collaboration} {Planck}),\ }\bibinfo {title} {{Planck 2018 results. VI. Cosmological parameters}},\ \href {https://doi.org/10.1051/0004-6361/201833910} {\bibfield  {journal} {Astron. Astrophys.\ }\textbf {\bibinfo {volume} {641}},\ \bibinfo {pages} {A6} (\bibinfo {year} {2020})},\ \bibinfo {note} {[Erratum: Astron.Astrophys. 652, C4 (2021)]}\BibitemShut {NoStop}%
\bibitem [{\citenamefont {Ade}\ {\it et~al.}(2018)\citenamefont {Ade} {\it et~al.}}]{BICEP2:2018kqh}%
  \BibitemOpen
  \bibfield  {author} {\bibinfo {author} {\bibfnamefont {P.~A.~R.}\ \bibnamefont {Ade}} {\it et~al.} (\bibinfo {collaboration} {BICEP2, Keck Array}),\ }\bibinfo {title} {{BICEP2 / Keck Array x: Constraints on Primordial Gravitational Waves using Planck, WMAP, and New BICEP2/Keck Observations through the 2015 Season}},\ \href {https://doi.org/10.1103/PhysRevLett.121.221301} {\bibfield  {journal} {Phys. Rev. Lett.\ }\textbf {\bibinfo {volume} {121}},\ \bibinfo {pages} {221301} (\bibinfo {year} {2018})}\BibitemShut {NoStop}%
\bibitem [{\citenamefont {Ade}\ {\it et~al.}(2021)\citenamefont {Ade} {\it et~al.}}]{BICEP:2021xfz}%
  \BibitemOpen
  \bibfield  {author} {\bibinfo {author} {\bibfnamefont {P.~A.~R.}\ \bibnamefont {Ade}} {\it et~al.} (\bibinfo {collaboration} {BICEP, Keck}),\ }\bibinfo {title} {{Improved Constraints on Primordial Gravitational Waves using Planck, WMAP, and BICEP/Keck Observations through the 2018 Observing Season}},\ \href {https://doi.org/10.1103/PhysRevLett.127.151301} {\bibfield  {journal} {Phys. Rev. Lett.\ }\textbf {\bibinfo {volume} {127}},\ \bibinfo {pages} {151301} (\bibinfo {year} {2021})}\BibitemShut {NoStop}%
\bibitem [{\citenamefont {Calabrese}\ {\it et~al.}(2025)\citenamefont {Calabrese} {\it et~al.}}]{ACT:2025tim}%
  \BibitemOpen
  \bibfield  {author} {\bibinfo {author} {\bibfnamefont {E.}~\bibnamefont {Calabrese}} {\it et~al.} (\bibinfo {collaboration} {ACT}),\ }\bibinfo {title} {{The Atacama Cosmology Telescope: DR6 Constraints on Extended Cosmological Models}},\ \Eprint {https://arxiv.org/abs/2503.14454} {arXiv:2503.14454} \BibitemShut {NoStop}%
\bibitem [{\citenamefont {Louis}\ {\it et~al.}(2025)\citenamefont {Louis} {\it et~al.}}]{ACT:2025fju}%
  \BibitemOpen
  \bibfield  {author} {\bibinfo {author} {\bibfnamefont {T.}~\bibnamefont {Louis}} {\it et~al.} (\bibinfo {collaboration} {ACT}),\ }\bibinfo {title} {{The Atacama Cosmology Telescope: DR6 Power Spectra, Likelihoods and $\Lambda$CDM Parameters}},\ \Eprint {https://arxiv.org/abs/2503.14452} {arXiv:2503.14452} \BibitemShut {NoStop}%
\bibitem [{\citenamefont {Cai}\ {\it et~al.}(2021)\citenamefont {Cai}, \citenamefont {Chen},\ and\ \citenamefont {Fu}}]{Cai:2021wzd}%
  \BibitemOpen
  \bibfield  {author} {\bibinfo {author} {\bibfnamefont {R.-G.}\ \bibnamefont {Cai}}, \bibinfo {author} {\bibfnamefont {C.}~\bibnamefont {Chen}},\ and\ \bibinfo {author} {\bibfnamefont {C.}~\bibnamefont {Fu}},\ }\bibinfo {title} {{Primordial black holes and stochastic gravitational wave background from inflation with a noncanonical spectator field}},\ \href {https://doi.org/10.1103/PhysRevD.104.083537} {\bibfield  {journal} {Phys. Rev. D\ }\textbf {\bibinfo {volume} {104}},\ \bibinfo {pages} {083537} (\bibinfo {year} {2021})}\BibitemShut {NoStop}%
\bibitem [{\citenamefont {Peng}\ {\it et~al.}(2021)\citenamefont {Peng}, \citenamefont {Fu}, \citenamefont {Liu}, \citenamefont {Guo},\ and\ \citenamefont {Cai}}]{Peng:2021zon}%
  \BibitemOpen
  \bibfield  {author} {\bibinfo {author} {\bibfnamefont {Z.-Z.}\ \bibnamefont {Peng}}, \bibinfo {author} {\bibfnamefont {C.}~\bibnamefont {Fu}}, \bibinfo {author} {\bibfnamefont {J.}~\bibnamefont {Liu}}, \bibinfo {author} {\bibfnamefont {Z.-K.}\ \bibnamefont {Guo}},\ and\ \bibinfo {author} {\bibfnamefont {R.-G.}\ \bibnamefont {Cai}},\ }\bibinfo {title} {{Gravitational waves from resonant amplification of curvature perturbations during inflation}},\ \href {https://doi.org/10.1088/1475-7516/2021/10/050} {J. Cosmol. Astropart. Phys.\ \bibinfo {volume} {10}\bibfield  {year} {\bibinfo  {year} { (2021)}\ }\bibinfo  {pages} {050}}\BibitemShut {NoStop}%
\bibitem [{\citenamefont {Fu}\ and\ \citenamefont {Wang}(2023)}]{Fu:2022ypp}%
  \BibitemOpen
\bibfield  {pages} {  }\bibfield  {author} {\bibinfo {author} {\bibfnamefont {C.}~\bibnamefont {Fu}}\ and\ \bibinfo {author} {\bibfnamefont {S.-J.}\ \bibnamefont {Wang}},\ }\bibinfo {title} {{Primordial black holes and induced gravitational waves from double-pole inflation}},\ \href {https://doi.org/10.1088/1475-7516/2023/06/012} {J. Cosmol. Astropart. Phys.\ \bibinfo {volume} {06}\bibfield  {year} {\bibinfo  {year} { (2023)}\ }\bibinfo  {pages} {012}}\BibitemShut {NoStop}%
\bibitem [{\citenamefont {Dimopoulos}(2017)}]{Dimopoulos:2017ged}%
  \BibitemOpen
\bibfield  {pages} {  }\bibfield  {author} {\bibinfo {author} {\bibfnamefont {K.}~\bibnamefont {Dimopoulos}},\ }\bibinfo {title} {{Ultra slow-roll inflation demystified}},\ \href {https://doi.org/10.1016/j.physletb.2017.10.066} {\bibfield  {journal} {Phys. Lett. B\ }\textbf {\bibinfo {volume} {775}},\ \bibinfo {pages} {262} (\bibinfo {year} {2017})}\BibitemShut {NoStop}%
\bibitem [{\citenamefont {Chung}(2003)}]{Chung:1998bt}%
  \BibitemOpen
  \bibfield  {author} {\bibinfo {author} {\bibfnamefont {D.~J.~H.}\ \bibnamefont {Chung}},\ }\bibinfo {title} {{Classical Inflation Field Induced Creation of Superheavy Dark Matter}},\ \href {https://doi.org/10.1103/PhysRevD.67.083514} {\bibfield  {journal} {Phys. Rev. D\ }\textbf {\bibinfo {volume} {67}},\ \bibinfo {pages} {083514} (\bibinfo {year} {2003})}\BibitemShut {NoStop}%
\bibitem [{\citenamefont {Chung}\ {\it et~al.}(2019)\citenamefont {Chung}, \citenamefont {Kolb},\ and\ \citenamefont {Long}}]{Chung:2018ayg}%
  \BibitemOpen
  \bibfield  {author} {\bibinfo {author} {\bibfnamefont {D.~J.~H.}\ \bibnamefont {Chung}}, \bibinfo {author} {\bibfnamefont {E.~W.}\ \bibnamefont {Kolb}},\ and\ \bibinfo {author} {\bibfnamefont {A.~J.}\ \bibnamefont {Long}},\ }\bibinfo {title} {{Gravitational production of super-Hubble-mass particles: an analytic approach}},\ \href {https://doi.org/10.1007/JHEP01(2019)189} {J. High Energ. Phys.\ \bibinfo {volume} {01}\bibfield  {year} {\bibinfo  {year} { (2019)}\ }\bibinfo  {pages} {189}}\BibitemShut {NoStop}%
\bibitem [{\citenamefont {Ford}(2021)}]{Ford:2021syk}%
  \BibitemOpen
\bibfield  {pages} {  }\bibfield  {author} {\bibinfo {author} {\bibfnamefont {L.~H.}\ \bibnamefont {Ford}},\ }\bibinfo {title} {{Cosmological particle production: a review}},\ \href {https://doi.org/10.1088/1361-6633/ac1b23} {\bibfield  {journal} {Rept. Prog. Phys.\ }\textbf {\bibinfo {volume} {84}}\  in press (\bibinfo {year} {2021})}\BibitemShut {NoStop}%
\bibitem [{\citenamefont {Kolb}\ and\ \citenamefont {Long}(2024)}]{Kolb:2023ydq}%
  \BibitemOpen
  \bibfield  {author} {\bibinfo {author} {\bibfnamefont {E.~W.}\ \bibnamefont {Kolb}}\ and\ \bibinfo {author} {\bibfnamefont {A.~J.}\ \bibnamefont {Long}},\ }\bibinfo {title} {{Cosmological gravitational particle production and its implications for cosmological relics}},\ \href {https://doi.org/10.1103/RevModPhys.96.045005} {\bibfield  {journal} {Rev. Mod. Phys.\ }\textbf {\bibinfo {volume} {96}},\ \bibinfo {pages} {045005} (\bibinfo {year} {2024})}\BibitemShut {NoStop}%
\bibitem [{\citenamefont {Chung}\ {\it et~al.}(2000)\citenamefont {Chung}, \citenamefont {Kolb}, \citenamefont {Riotto},\ and\ \citenamefont {Tkachev}}]{Chung:1999ve}%
  \BibitemOpen
  \bibfield  {author} {\bibinfo {author} {\bibfnamefont {D.~J.~H.}\ \bibnamefont {Chung}}, \bibinfo {author} {\bibfnamefont {E.~W.}\ \bibnamefont {Kolb}}, \bibinfo {author} {\bibfnamefont {A.}~\bibnamefont {Riotto}},\ and\ \bibinfo {author} {\bibfnamefont {I.~I.}\ \bibnamefont {Tkachev}},\ }\bibinfo {title} {{Probing Planckian physics: Resonant production of particles during inflation and features in the primordial power spectrum}},\ \href {https://doi.org/10.1103/PhysRevD.62.043508} {\bibfield  {journal} {Phys. Rev. D\ }\textbf {\bibinfo {volume} {62}},\ \bibinfo {pages} {043508} (\bibinfo {year} {2000})}\BibitemShut {NoStop}%
\bibitem [{\citenamefont {Romano}\ and\ \citenamefont {Sasaki}(2008)}]{Romano:2008rr}%
  \BibitemOpen
  \bibfield  {author} {\bibinfo {author} {\bibfnamefont {A.~E.}\ \bibnamefont {Romano}}\ and\ \bibinfo {author} {\bibfnamefont {M.}~\bibnamefont {Sasaki}},\ }\bibinfo {title} {{Effects of particle production during inflation}},\ \href {https://doi.org/10.1103/PhysRevD.78.103522} {\bibfield  {journal} {Phys. Rev. D\ }\textbf {\bibinfo {volume} {78}},\ \bibinfo {pages} {103522} (\bibinfo {year} {2008})}\BibitemShut {NoStop}%
\bibitem [{\citenamefont {Battefeld}\ and\ \citenamefont {Battefeld}(2010)}]{Battefeld:2010sw}%
  \BibitemOpen
  \bibfield  {author} {\bibinfo {author} {\bibfnamefont {D.}~\bibnamefont {Battefeld}}\ and\ \bibinfo {author} {\bibfnamefont {T.}~\bibnamefont {Battefeld}},\ }\bibinfo {title} {{A Terminal Velocity on the Landscape: Particle Production near Extra Species Loci in Higher Dimensions}},\ \href {https://doi.org/10.1007/JHEP07(2010)063} {J. High Energ. Phys.\ \bibinfo {volume} {07}\bibfield  {year} {\bibinfo  {year} { (2010)}\ }\bibinfo  {pages} {063}}\BibitemShut {NoStop}%
\bibitem [{\citenamefont {Cook}\ and\ \citenamefont {Sorbo}(2012)}]{Cook:2011hg}%
  \BibitemOpen
\bibfield  {pages} {  }\bibfield  {author} {\bibinfo {author} {\bibfnamefont {J.~L.}\ \bibnamefont {Cook}}\ and\ \bibinfo {author} {\bibfnamefont {L.}~\bibnamefont {Sorbo}},\ }\bibinfo {title} {{Particle production during inflation and gravitational waves detectable by ground-based interferometers}},\ \href {https://doi.org/10.1103/PhysRevD.85.023534} {\bibfield  {journal} {Phys. Rev. D\ }\textbf {\bibinfo {volume} {85}},\ \bibinfo {pages} {023534} (\bibinfo {year} {2012})},\ \bibinfo {note} {[Erratum: Phys.Rev.D 86, 069901 (2012)]}\BibitemShut {NoStop}%
\bibitem [{\citenamefont {Fedderke}\ {\it et~al.}(2015)\citenamefont {Fedderke}, \citenamefont {Kolb},\ and\ \citenamefont {Wyman}}]{Fedderke:2014ura}%
  \BibitemOpen
  \bibfield  {author} {\bibinfo {author} {\bibfnamefont {M.~A.}\ \bibnamefont {Fedderke}}, \bibinfo {author} {\bibfnamefont {E.~W.}\ \bibnamefont {Kolb}},\ and\ \bibinfo {author} {\bibfnamefont {M.}~\bibnamefont {Wyman}},\ }\bibinfo {title} {{Irruption of massive particle species during inflation}},\ \href {https://doi.org/10.1103/PhysRevD.91.063505} {\bibfield  {journal} {Phys. Rev. D\ }\textbf {\bibinfo {volume} {91}},\ \bibinfo {pages} {063505} (\bibinfo {year} {2015})}\BibitemShut {NoStop}%
\bibitem [{\citenamefont {Kim}\ {\it et~al.}(2021)\citenamefont {Kim}, \citenamefont {Kumar}, \citenamefont {Martin},\ and\ \citenamefont {Tsai}}]{Kim:2021ida}%
  \BibitemOpen
  \bibfield  {author} {\bibinfo {author} {\bibfnamefont {J.~H.}\ \bibnamefont {Kim}}, \bibinfo {author} {\bibfnamefont {S.}~\bibnamefont {Kumar}}, \bibinfo {author} {\bibfnamefont {A.}~\bibnamefont {Martin}},\ and\ \bibinfo {author} {\bibfnamefont {Y.}~\bibnamefont {Tsai}},\ }\bibinfo {title} {{Cosmological particle production and pairwise hotspots on the CMB}},\ \href {https://doi.org/10.1007/JHEP11(2021)158} {J. High Energ. Phys.\ \bibinfo {volume} {11}\bibfield  {year} {\bibinfo  {year} { (2021)}\ }\bibinfo  {pages} {158}}\BibitemShut {NoStop}%
\bibitem [{\citenamefont {Li}\ {\it et~al.}(2019)\citenamefont {Li}, \citenamefont {Nakama}, \citenamefont {Sou}, \citenamefont {Wang},\ and\ \citenamefont {Zhou}}]{Li:2019ves}%
  \BibitemOpen
\bibfield  {pages} {  }\bibfield  {author} {\bibinfo {author} {\bibfnamefont {L.}~\bibnamefont {Li}}, \bibinfo {author} {\bibfnamefont {T.}~\bibnamefont {Nakama}}, \bibinfo {author} {\bibfnamefont {C.~M.}\ \bibnamefont {Sou}}, \bibinfo {author} {\bibfnamefont {Y.}~\bibnamefont {Wang}},\ and\ \bibinfo {author} {\bibfnamefont {S.}~\bibnamefont {Zhou}},\ }\bibinfo {title} {{Gravitational Production of Superheavy Dark Matter and Associated Cosmological Signatures}},\ \href {https://doi.org/10.1007/JHEP07(2019)067} {J. High Energ. Phys.\ \bibinfo {volume} {07}\bibfield  {year} {\bibinfo  {year} { (2019)}\ }\bibinfo  {pages} {067}}\BibitemShut {NoStop}%
\bibitem [{\citenamefont {Clery}\ {\it et~al.}(2022)\citenamefont {Clery}, \citenamefont {Mambrini}, \citenamefont {Olive}, \citenamefont {Shkerin},\ and\ \citenamefont {Verner}}]{Clery:2022wib}%
  \BibitemOpen
\bibfield  {pages} {  }\bibfield  {author} {\bibinfo {author} {\bibfnamefont {S.}~\bibnamefont {Clery}}, \bibinfo {author} {\bibfnamefont {Y.}~\bibnamefont {Mambrini}}, \bibinfo {author} {\bibfnamefont {K.~A.}\ \bibnamefont {Olive}}, \bibinfo {author} {\bibfnamefont {A.}~\bibnamefont {Shkerin}},\ and\ \bibinfo {author} {\bibfnamefont {S.}~\bibnamefont {Verner}},\ }\bibinfo {title} {{Gravitational portals with nonminimal couplings}},\ \href {https://doi.org/10.1103/PhysRevD.105.095042} {\bibfield  {journal} {Phys. Rev. D\ }\textbf {\bibinfo {volume} {105}},\ \bibinfo {pages} {095042} (\bibinfo {year} {2022})}\BibitemShut {NoStop}%
\bibitem [{\citenamefont {Yu}\ {\it et~al.}(2023)\citenamefont {Yu}, \citenamefont {Fu},\ and\ \citenamefont {Guo}}]{Yu:2023ity}%
  \BibitemOpen
  \bibfield  {author} {\bibinfo {author} {\bibfnamefont {Z.}~\bibnamefont {Yu}}, \bibinfo {author} {\bibfnamefont {C.}~\bibnamefont {Fu}},\ and\ \bibinfo {author} {\bibfnamefont {Z.-K.}\ \bibnamefont {Guo}},\ }\bibinfo {title} {{Particle production during inflation with a nonminimally coupled spectator scalar field}},\ \href {https://doi.org/10.1103/PhysRevD.108.123509} {\bibfield  {journal} {Phys. Rev. D\ }\textbf {\bibinfo {volume} {108}},\ \bibinfo {pages} {123509} (\bibinfo {year} {2023})}\BibitemShut {NoStop}%
\bibitem [{\citenamefont {Capanelli}\ {\it et~al.}(2024)\citenamefont {Capanelli}, \citenamefont {Jenks}, \citenamefont {Kolb},\ and\ \citenamefont {McDonough}}]{Capanelli:2024rlk}%
  \BibitemOpen
  \bibfield  {author} {\bibinfo {author} {\bibfnamefont {C.}~\bibnamefont {Capanelli}}, \bibinfo {author} {\bibfnamefont {L.}~\bibnamefont {Jenks}}, \bibinfo {author} {\bibfnamefont {E.~W.}\ \bibnamefont {Kolb}},\ and\ \bibinfo {author} {\bibfnamefont {E.}~\bibnamefont {McDonough}},\ }\bibinfo {title} {{Gravitational production of completely dark photons with nonminimal couplings to gravity}},\ \href {https://doi.org/10.1007/JHEP09(2024)071} {J. High Energ. Phys.\ \bibinfo {volume} {09}\bibfield  {year} {\bibinfo  {year} { (2024)}\ }\bibinfo  {pages} {071}}\BibitemShut {NoStop}%
\bibitem [{\citenamefont {He}\ {\it et~al.}(2025)\citenamefont {He}, \citenamefont {Fu}, \citenamefont {Zhang},\ and\ \citenamefont {Guo}}]{He:2024bno}%
  \BibitemOpen
\bibfield  {pages} {  }\bibfield  {author} {\bibinfo {author} {\bibfnamefont {J.-F.}\ \bibnamefont {He}}, \bibinfo {author} {\bibfnamefont {C.}~\bibnamefont {Fu}}, \bibinfo {author} {\bibfnamefont {K.-G.}\ \bibnamefont {Zhang}},\ and\ \bibinfo {author} {\bibfnamefont {Z.-K.}\ \bibnamefont {Guo}},\ }\bibinfo {title} {{Gravitational waves from a gauge field nonminimally coupled to gravity}},\ \href {https://doi.org/10.1103/PhysRevD.111.023536} {\bibfield  {journal} {Phys. Rev. D\ }\textbf {\bibinfo {volume} {111}},\ \bibinfo {pages} {023536} (\bibinfo {year} {2025})}\BibitemShut {NoStop}%
\bibitem [{\citenamefont {Danzmann}(1997)}]{Danzmann:1997hm}%
  \BibitemOpen
  \bibfield  {author} {\bibinfo {author} {\bibfnamefont {K.}~\bibnamefont {Danzmann}},\ }\bibinfo {title} {{LISA: An ESA cornerstone mission for a gravitational wave observatory}},\ \href {https://doi.org/10.1088/0264-9381/14/6/002} {\bibfield  {journal} {Classical Quantum Gravity\ }\textbf {\bibinfo {volume} {14}},\ \bibinfo {pages} {1399} (\bibinfo {year} {1997})}\BibitemShut {NoStop}%
\bibitem [{\citenamefont {Amaro-Seoane}\ {\it et~al.}(2017)\citenamefont {Amaro-Seoane} {\it et~al.}}]{LISA:2017pwj}%
  \BibitemOpen
  \bibfield  {author} {\bibinfo {author} {\bibfnamefont {P.}~\bibnamefont {Amaro-Seoane}} {\it et~al.} (\bibinfo {collaboration} {LISA}),\ }\bibinfo {title} {{Laser Interferometer Space Antenna}},\ \Eprint {https://arxiv.org/abs/1702.00786} {arXiv:1702.00786} \BibitemShut {NoStop}%
\bibitem [{\citenamefont {Luo}\ {\it et~al.}(2016)\citenamefont {Luo} {\it et~al.}}]{Luo:2015ght}%
  \BibitemOpen
  \bibfield  {author} {\bibinfo {author} {\bibfnamefont {J.}~\bibnamefont {Luo}} {\it et~al.} (\bibinfo {collaboration} {TianQin}),\ }\bibinfo {title} {{TianQin: a space-borne gravitational wave detector}},\ \href {https://doi.org/10.1088/0264-9381/33/3/035010} {\bibfield  {journal} {Classical Quantum Gravity\ }\textbf {\bibinfo {volume} {33}},\ \bibinfo {pages} {035010} (\bibinfo {year} {2016})}\BibitemShut {NoStop}%
\bibitem [{\citenamefont {Gong}\ {\it et~al.}(2021)\citenamefont {Gong}, \citenamefont {Luo},\ and\ \citenamefont {Wang}}]{Gong:2021gvw}%
  \BibitemOpen
  \bibfield  {author} {\bibinfo {author} {\bibfnamefont {Y.}~\bibnamefont {Gong}}, \bibinfo {author} {\bibfnamefont {J.}~\bibnamefont {Luo}},\ and\ \bibinfo {author} {\bibfnamefont {B.}~\bibnamefont {Wang}},\ }\bibinfo {title} {{Concepts and status of Chinese space gravitational wave detection projects}},\ \href {https://doi.org/10.1038/s41550-021-01480-3} {\bibfield  {journal} {Nat. Astron.\ }\textbf {\bibinfo {volume} {5}},\ \bibinfo {pages} {881} (\bibinfo {year} {2021})}\BibitemShut {NoStop}%
\bibitem [{\citenamefont {Hu}\ and\ \citenamefont {Wu}(2017)}]{Hu:2017mde}%
  \BibitemOpen
  \bibfield  {author} {\bibinfo {author} {\bibfnamefont {W.-R.}\ \bibnamefont {Hu}}\ and\ \bibinfo {author} {\bibfnamefont {Y.-L.}\ \bibnamefont {Wu}},\ }\bibinfo {title} {{The Taiji Program in Space for gravitational wave physics and the nature of gravity}},\ \href {https://doi.org/10.1093/nsr/nwx116} {\bibfield  {journal} {Natl. Sci. Rev.\ }\textbf {\bibinfo {volume} {4}},\ \bibinfo {pages} {685} (\bibinfo {year} {2017})}\BibitemShut {NoStop}%
\bibitem [{\citenamefont {Yang}\ {\it et~al.}(2025)\citenamefont {Yang}, \citenamefont {Yu},\ and\ \citenamefont {Cao}}]{Yang:2025zap}%
  \BibitemOpen
  \bibfield  {author} {\bibinfo {author} {\bibfnamefont {X.}~\bibnamefont {Yang}}, \bibinfo {author} {\bibfnamefont {Z.}~\bibnamefont {Yu}},\ and\ \bibinfo {author} {\bibfnamefont {Z.}~\bibnamefont {Cao}},\ }\bibinfo {title} {{Impact of particle production during inflation on the CMB detection}},\ \Eprint {https://arxiv.org/abs/2501.08592} {arXiv:2501.08592} \BibitemShut {NoStop}%
\bibitem [{\citenamefont {Barnaby}\ {\it et~al.}(2009)\citenamefont {Barnaby}, \citenamefont {Huang}, \citenamefont {Kofman},\ and\ \citenamefont {Pogosyan}}]{Barnaby:2009mc}%
  \BibitemOpen
  \bibfield  {author} {\bibinfo {author} {\bibfnamefont {N.}~\bibnamefont {Barnaby}}, \bibinfo {author} {\bibfnamefont {Z.}~\bibnamefont {Huang}}, \bibinfo {author} {\bibfnamefont {L.}~\bibnamefont {Kofman}},\ and\ \bibinfo {author} {\bibfnamefont {D.}~\bibnamefont {Pogosyan}},\ }\bibinfo {title} {{Cosmological Fluctuations from Infra-Red Cascading During Inflation}},\ \href {https://doi.org/10.1103/PhysRevD.80.043501} {\bibfield  {journal} {Phys. Rev. D\ }\textbf {\bibinfo {volume} {80}},\ \bibinfo {pages} {043501} (\bibinfo {year} {2009})}\BibitemShut {NoStop}%
\bibitem [{\citenamefont {Pearce}\ {\it et~al.}(2017)\citenamefont {Pearce}, \citenamefont {Peloso},\ and\ \citenamefont {Sorbo}}]{Pearce:2017bdc}%
  \BibitemOpen
  \bibfield  {author} {\bibinfo {author} {\bibfnamefont {L.}~\bibnamefont {Pearce}}, \bibinfo {author} {\bibfnamefont {M.}~\bibnamefont {Peloso}},\ and\ \bibinfo {author} {\bibfnamefont {L.}~\bibnamefont {Sorbo}},\ }\bibinfo {title} {{Resonant particle production during inflation: a full analytical study}},\ \href {https://doi.org/10.1088/1475-7516/2017/05/054} {J. Cosmol. Astropart. Phys.\ \bibinfo {volume} {05}\bibfield  {year} {\bibinfo  {year} { (2017)}\ }\bibinfo  {pages} {054}}\BibitemShut {NoStop}%
\bibitem [{\citenamefont {Fumagalli}\ {\it et~al.}(2024)\citenamefont {Fumagalli}, \citenamefont {Bhattacharya}, \citenamefont {Peloso}, \citenamefont {Renaux-Petel},\ and\ \citenamefont {Witkowski}}]{Fumagalli:2023loc}%
  \BibitemOpen
\bibfield  {pages} {  }\bibfield  {author} {\bibinfo {author} {\bibfnamefont {J.}~\bibnamefont {Fumagalli}}, \bibinfo {author} {\bibfnamefont {S.}~\bibnamefont {Bhattacharya}}, \bibinfo {author} {\bibfnamefont {M.}~\bibnamefont {Peloso}}, \bibinfo {author} {\bibfnamefont {S.}~\bibnamefont {Renaux-Petel}},\ and\ \bibinfo {author} {\bibfnamefont {L.~T.}\ \bibnamefont {Witkowski}},\ }\bibinfo {title} {{One-loop infrared rescattering by enhanced scalar fluctuations during inflation}},\ \href {https://doi.org/10.1088/1475-7516/2024/04/029} {J. Cosmol. Astropart. Phys.\ \bibinfo {volume} {04}\bibfield  {year} {\bibinfo  {year} { (2024)}\ }\bibinfo  {pages} {029}}\BibitemShut {NoStop}%
\bibitem [{\citenamefont {Chung}\ {\it et~al.}(2005)\citenamefont {Chung}, \citenamefont {Kolb}, \citenamefont {Riotto},\ and\ \citenamefont {Senatore}}]{Chung:2004nh}%
  \BibitemOpen
\bibfield  {pages} {  }\bibfield  {author} {\bibinfo {author} {\bibfnamefont {D.~J.~H.}\ \bibnamefont {Chung}}, \bibinfo {author} {\bibfnamefont {E.~W.}\ \bibnamefont {Kolb}}, \bibinfo {author} {\bibfnamefont {A.}~\bibnamefont {Riotto}},\ and\ \bibinfo {author} {\bibfnamefont {L.}~\bibnamefont {Senatore}},\ }\bibinfo {title} {{Isocurvature constraints on gravitationally produced superheavy dark matter}},\ \href {https://doi.org/10.1103/PhysRevD.72.023511} {\bibfield  {journal} {Phys. Rev. D\ }\textbf {\bibinfo {volume} {72}},\ \bibinfo {pages} {023511} (\bibinfo {year} {2005})}\BibitemShut {NoStop}%
\bibitem [{\citenamefont {Chen}\ {\it et~al.}(2007)\citenamefont {Chen}, \citenamefont {Huang}, \citenamefont {Kachru},\ and\ \citenamefont {Shiu}}]{Chen:2006nt}%
  \BibitemOpen
  \bibfield  {author} {\bibinfo {author} {\bibfnamefont {X.}~\bibnamefont {Chen}}, \bibinfo {author} {\bibfnamefont {M.-x.}\ \bibnamefont {Huang}}, \bibinfo {author} {\bibfnamefont {S.}~\bibnamefont {Kachru}},\ and\ \bibinfo {author} {\bibfnamefont {G.}~\bibnamefont {Shiu}},\ }\bibinfo {title} {{Observational signatures and non-Gaussianities of general single field inflation}},\ \href {https://doi.org/10.1088/1475-7516/2007/01/002} {J. Cosmol. Astropart. Phys.\ \bibinfo {volume} {01}\bibfield  {year} {\bibinfo  {year} { (2007)}\ }\bibinfo  {pages} {002}}\BibitemShut {NoStop}%
\bibitem [{\citenamefont {Caravano}\ {\it et~al.}(2025{\natexlab{a}})\citenamefont {Caravano}, \citenamefont {Franciolini},\ and\ \citenamefont {Renaux-Petel}}]{Caravano:2024moy}%
  \BibitemOpen
\bibfield  {pages} {  }\bibfield  {author} {\bibinfo {author} {\bibfnamefont {A.}~\bibnamefont {Caravano}}, \bibinfo {author} {\bibfnamefont {G.}~\bibnamefont {Franciolini}},\ and\ \bibinfo {author} {\bibfnamefont {S.}~\bibnamefont {Renaux-Petel}},\ }\bibinfo {title} {{Ultraslow-roll inflation on the lattice: Backreaction and nonlinear effects}},\ \href {https://doi.org/10.1103/PhysRevD.111.063518} {\bibfield  {journal} {Phys. Rev. D\ }\textbf {\bibinfo {volume} {111}},\ \bibinfo {pages} {063518} (\bibinfo {year} {2025}{\natexlab{a}})}\BibitemShut {NoStop}%
\bibitem [{\citenamefont {Caravano}\ {\it et~al.}(2025{\natexlab{b}})\citenamefont {Caravano}, \citenamefont {Franciolini},\ and\ \citenamefont {Renaux-Petel}}]{Caravano:2025diq}%
  \BibitemOpen
  \bibfield  {author} {\bibinfo {author} {\bibfnamefont {A.}~\bibnamefont {Caravano}}, \bibinfo {author} {\bibfnamefont {G.}~\bibnamefont {Franciolini}},\ and\ \bibinfo {author} {\bibfnamefont {S.}~\bibnamefont {Renaux-Petel}},\ }\bibinfo {title} {{Ultra-Slow-Roll Inflation on the Lattice II: Nonperturbative Curvature Perturbation}},\ \Eprint {https://arxiv.org/abs/2506.11795} {arXiv:2506.11795} \BibitemShut {NoStop}%
\end{thebibliography}%
\end{document}